\documentclass[prb,twocolumn,aps,showpacs,amsmath,amssymb,superscriptaddress]{revtex4-1}

\usepackage{color}
\usepackage{bm}
\usepackage{graphicx}
\usepackage{braket}

\usepackage{color}

\newcommand{\bea}{\begin{eqnarray}}
\newcommand{\eea}{\end{eqnarray}}
\newcommand{\la}{\label}
\newcommand{\be}{\begin{equation}}
\newcommand{\ee}{\end{equation}}

\newcommand{\sgn}{\,\mbox{sgn}}



\begin{document}
\title{Many-body localization in incommensurate models with a mobility edge }

\author{Dong-Ling Deng} 
\affiliation{Condensed Matter Theory Center and Joint Quantum Institute, Department
of Physics, University of Maryland, College Park, MD 20742-4111, USA}
\author{Sriram Ganeshan}
\affiliation{Simons Center for Geometry and Physics, Stony Brook, NY 11794, USA}
\author{Xiaopeng Li}
\affiliation{Condensed Matter Theory Center and Joint Quantum Institute, Department
of Physics, University of Maryland, College Park, MD 20742-4111, USA}
\affiliation{Department of Physics, Fudan University, Shanghai, China} 
\author{Ranjan Modak} 
\author{Subroto Mukerjee}
\affiliation{Department of Physics, Indian Institute of Science, Bangalore 560 012, India}
\author{J. H. Pixley}
\affiliation{Condensed Matter Theory Center and Joint Quantum Institute, Department
of Physics, University of Maryland, College Park, MD 20742-4111, USA}

\begin{abstract}
We review the physics of many-body localization in models with incommensurate potentials. In particular, we consider one-dimensional quasiperiodic models with single-particle mobility edges. A conventional perspective suggests that delocalized states act as a thermalizing bath for the localized states in the presence of of interactions. However, contrary to this intuition there is evidence that such systems can display non-ergodicity. This is in part due to the fact that the delocalized states do not have any kind of protection due to  symmetry or topology and are thus susceptible to localization. A study of such incommensurate models, in the non-interacting limit, shows that they admit extended, partially extended, and fully localized many-body states. Non-interacting incommensurate models cannot thermalize dynamically and remain localized upon the introduction of interactions. In particular, for a certain range of energy, the system can host a non-ergodic extended (i.e. metallic) phase in which the energy eigenstates violate the eigenstate thermalization hypothesis (ETH) but the entanglement entropy obeys volume-law scaling.  The level statistics and entanglement growth also indicate the lack of ergodicity in these models. The phenomenon of localization and non-ergodicity in a system with interactions despite the presence of single-particle delocalized states is closely related to the so-called ``many-body proximity effect'' and can also be observed in models with disorder coupled to systems with delocalized degrees of freedom. Many-body localization in systems with incommensurate potentials (without single-particle mobility edges) have been  realized experimentally, and we show how this can be modified to study the the effects of such mobility edges. Demonstrating the failure of thermalization in the presence of a single-particle mobility edge in the thermodynamic limit would indicate a more robust violation of the ETH.	
\end{abstract}
\maketitle

\section{Introduction} 
Isolated quantum systems have become a model setting to understand the physics of thermalization. 
Due to unitary time evolution, it is not obvious how a quantum state will reach thermal equilibrium. 
If we consider a subsystem ($A$) of the entire system it is conceivable how $A$ can reach thermal equilibrium; if the remaining degrees of freedom are sufficiently entangled with $A$, they can act like a ``bath'' that efficiently thermalizes the subsystem $A$ (see Refs.~\onlinecite{2015_NH_review,2015_Altman_review} for recent reviews). The theoretical underpinnings for how such an isolated quantum system can reach thermal equilibrium have been put forth as the eigenstate thermalization hypothesis (ETH) (Refs.~\onlinecite{Deutsch_1991_PRA,Srednicki_1994_PRL, 2008_Rigol_Nature}). However, if interacting many-body systems are subjected to strong disorder they can become many-body localized, where ETH fails and random initial states \emph{never} relax. There is now a tremendous amount of effort being devoted to understand many-body localization (MBL), with mounting theoretical evidence for the existence of the MBL phase, through perturbative analysis~\cite{Basko_2006_AP}, renormalization group approaches~\cite{2015_Potter_PRX, 2015_Vosk_PRX,Zhang_2016_PRB,2015_Altman_review}, exact numerical studies~\cite{Oganesyan_2007_PRB, 2008_Marko_PRB, Pal_2010_PRB,Bardarson_2014_PRL,Luitz_2015_PRB}, and a mathematical proof~\cite{Imbrie_2016_JSP}. More recently, experiments in ultracold atomic gases and trapped-ion simulators have reported observations of MBL with initial ``high temperature'' states that never relax for the entire experimental run time~\cite{Schreiber-Science-2015, Kondov_2015_PRL,Smith-2016-NaturePhysics,Bordia-2016-PRL,Choi-2016-Science,Bordia-2016-arXiv,Luschen-2016-arXiv,Zhang-2016-arXiv}.

The most natural starting point to study MBL is by considering a non-interacting system with every eigenstate of the Hamiltonian being Anderson localized~\cite{Anderson_1958_PR} and then ``turning on'' weak interactions to see if this phase remains stable.
Thus, the starting point has exponentially decaying single-particle eigenstates and an energy spectrum that is dense with no gap. 
In the highly excited states of a thermal system the entanglement entropy scales with the volume of the subsystem~\cite{Kim_PRL_2013} (i.e. volume-law scaling) and the eigenvalues have a non-zero level repulsion~\cite{Mukerjee_2006_PRB}. Whereas, in the MBL phase at large disorder, the entanglement entropy scales with the boundary of the subsystem (i.e. an area law scaling)~\cite{2013_Bauer_JSM} and there is a complete absence of level repulsion giving rise to Poisson level statistics~\cite{Oganesyan_2007_PRB}. In the MBL phase, statistical fluctuations of nearby energy eigenstates are so large that ETH completely fails. For a Hamiltonian that is in the MBL phase, a global quench cannot induce any DC transport but instead leads to de-phasing, which produces an entanglement entropy that grows logarithmically in time~\cite{Bardarson_2012_PRL,Serbyn_2013_PRL,Nanduri_2014_PRB}.

It is also possible for MBL to occur in a system with no random disorder. In single-particle Hamiltonians, it is well known that non-random incommensurate potentials can induce Anderson localization~\cite{Aubry_1980_AIPS,Grempel_1982_PRL,Prange_1985_book,Sankar_1986_PRL,Thouless_1988_PRL,Sankar_1988_PRL,Biddle_2009_PRA,Biddle_2011_PRB}. 
Interestingly, in the presence of interactions, this can also lead to MBL~\cite{Iyer_2013_PRB}. However, it is important to emphasize that incommensurate potential driven localization is inherently different from that due to random disorder. Instead of suppressing single-particle tunneling through large potential deviations in nearby sites, it is the multifractal gap structure that ultimately leads to localization. Thus, there are various qualitative distinctions between these two phenomena. For example, incommensurate models lack a random-matrix-theory~\cite{Brody_1981_RMP} description and there are no rigorous bounds on the critical exponents concerning the transition~\cite{Chayes_1986_PRL}. However, there are many physical features in common, such as their wave functions being multifractal at the transition and exponentially localized in each respective localized phase.

In one and two dimensions (1D, 2D), all eigenstates of a generic non-interacting system will be localized for an infinitesimal disorder strength~\cite{Abrahams_1979_PRL} (for certain symmetry classes), thus there is no localization transition at a finite disorder strength. In contrast, a rather useful property of incommensurate models is the access to a localization-delocalization transition in 1D. In ultra-cold atom experiments, incommensurate potentials can be engineered straightforwardly and have led to a number of interesting studies of localization. For the Aubry-Andre (AA) model, either all states are delocalized or localized depending on the relative strength of the incommensurate potential. By ``tweaking'' the incommensurate potential in the AA model it is possible to introduce a single-particle mobility edge into the model~\cite{2010_Biddle_PRL,Sriram_2015_PRL}, which separates localized and delocalized eigenstates in energy. In disordered models, single-particle mobility edges are common in 3D~\cite{Lee_1985_RMP,Evers_2008_RMP}. Thus, incommensurate models open the door to effectively simulate some properties of Anderson localization that occur generically in higher dimensions. This is particularly tantalizing for numerical studies, where the many-body problem can only be handled with small system sizes due to the exponentially large Hilbert space, and as a result, any numerical studies in dimensions greater than one are at present futile (this may be overcome with new numerical approaches such as Refs.~\onlinecite{2015_Yu_Pekker_arXiv, 2015_Khemani_Pollmann_arXiv, Inglis_2016_PRL}). In this context for MBL, incommensurate models offer a unique opportunity to probe physics that is out of reach using random disordered potentials. 

The stability of a fully MBL phase can naturally be captured through an extensive number of local integrals of motion (LIOM)~\cite{2013_Serbyn_PRL,Huse_2014_PRB,Chandran_2015_PRB,Ros_2015_NPB}. However, this phenomenological description
is much more nuanced when starting from models that don't necessarily have all single-particle states localized. If instead, we consider a generalized AA (GAA) model that has a single-particle mobility edge at a finite filling, a non-interacting wave function of the model consists of a Slater determinant of \emph{both} delocalized and localized orbitals~\cite{Li_2016_PRB, Modak_2016_arxiv}. If we now ``switch on'' interactions in the model, these eigenstates will interact in a rather complicated way and the interaction between extended and localized orbitals could delocalize all of the states resulting in only a thermal phase. However, there is also the possibility that the localized orbitals can act like an incommensurate potential that localizes all of the delocalized orbitals creating a stable MBL phase~\cite{Li_2015_PRL,Modak_2015_PRL}. The latter phenomenon has been dubbed the many-body localization proximity effect~\cite{2015_Rahul_PRB}, to describe the general scenario of strongly localized orbitals inducing localization, which may also occur in distinct settings such as ladder models where there is a \emph{spatial} separation between localized and delocalized states~\cite{Hyatt_2016_arXiv}. There is also a third, rather unconventional possibility, where the localization and thermalization transitions become distinct transitions and as a result the presence of a phase intervening between the thermal and MBL phase~\cite{Li_2015_PRL,Modak_2015_PRL}. The so-called non-ergodic metal will violate ETH but contain extended states.

In the following topical review we will discuss in detail  
the physical similarities and differences between single-particle localization driven by either incommensurate potentials or random disorder and what their implications are on the MBL phase. Following along these lines we will discuss all of the physical possibilities afforded to studying incommensurate models, focusing heavily on MBL when starting from a model with a single-particle mobility edge. We will review the existing work on the existence of MBL that results from interacting localized and delocalized orbitals in both incommensurate and ladder models. We will also discuss the experimental progress of MBL using ultracold atomic gases that engineer an incommensurate lattice potential.

\section{Single-particle localization in quasiperiodic potentials}
\subsection{Mobility edge in quasiperiodic potentials}
We begin by first reviewing the duality of the AA model. Aubry and Andre~\cite{Aubry_1980_AIPS} were the first to demonstrate the existence of a localization transition in a 1D quasiperiodic lattice model. Below, we present the derivation following the original Aubry-Andre paper~\cite{Aubry_1980_AIPS}, which will set up the discussion for the more general setting. The existence of localization transition is attributed to a self-dual symmetry. Consider the following Schr\"odinger equation,
\begin{align}
	t(\psi_{n+1}+\psi_{n-1})+2\lambda \cos(2\pi n b+\phi)\psi_n=E\psi_n,
	\label{eq:aa}
\end{align}
where $b$ is an irrational number that makes the on-site potential incommensurate with respect to the lattice periodicity. We do a transformation $\psi_n=e^{ikn}\sum_{m=-\infty}^{\infty}e^{i m(2 n\pi b +\phi)}f_m$ to rewrite Eq.~(\ref{eq:aa}) in terms of $f$, 
\begin{align}
	\lambda(f_{m+1}+f_{m-1})+2t \cos(2\pi m b+k)f_m=E f_m. 
	\label{eq:aadual}
\end{align}
This is the same equation as before, if we set $\lambda\rightarrow t$, $\phi\rightarrow k$ and $\{f\}\equiv\{\psi\}$. A localized solution in the $f$ space is a delocalized solution in the  $\psi$ space and vice versa. This places the localization transition at $\lambda_c/t=1$. However, the duality symmetry itself a priori does not indicate which states are localized or delocalized. This duality manifests in the density of states expression as,
\begin{align}
\rho_{\lambda/t}(E/t)=\rho_{t/\lambda}(E/\lambda),
\end{align}
which is an energy \emph{independent} duality. Using the Thouless formula $\gamma(E)=\int^{\infty}_{-\infty} \ln |E-E'|\rho(E')dE'$ in combination with the duality allows one to
 deduce the localization length $\xi$ of all the eigenstates,
 \begin{align}
 	\xi=\frac{1}{\ln(\lambda/t)}.
	\label{eqn:xi}
 \end{align}

Aubry and Andre in their original paper used numerical arguments to rule out mobility edges in self-dual models. However, in recent work, two of the authors of this review showed that mobility edge can exist for a generalized duality symmetry in a class of deformed AA models.  This generalized duality symmetry is a symmetry of the full Schr\"odinger equation. The GAA model can be written as,
\begin{equation}
 t (\psi_{n-1}+\psi_{n+1})+V_n(\alpha,\phi)\psi_n=E\psi_n.\label{modelI}
 \end{equation}
 The on-site potential, $V_{n}(\alpha, \phi)$, is characterized by the deformation parameter $\alpha$, on-site modulation strength $\lambda$, period $1/b$ and the phase parameter $\phi$. For a quasiperiodic modulation, we set $b$ to be irrational. The first family of models we consider are specified by an on-site potential
 \begin{equation}
 V_n(\alpha,\phi)=2\lambda\frac{\cos (2\pi n b+\phi)}{1-\alpha\cos (2\pi n b+\phi)}.\la{onsite}
 \end{equation}
 This on-site potential is a smooth function of $\alpha$ in the open interval $\alpha \in (-1,1)$. We recover the AA model for $\alpha=0$. 
\begin{figure}
\centering
\includegraphics[scale=0.52]{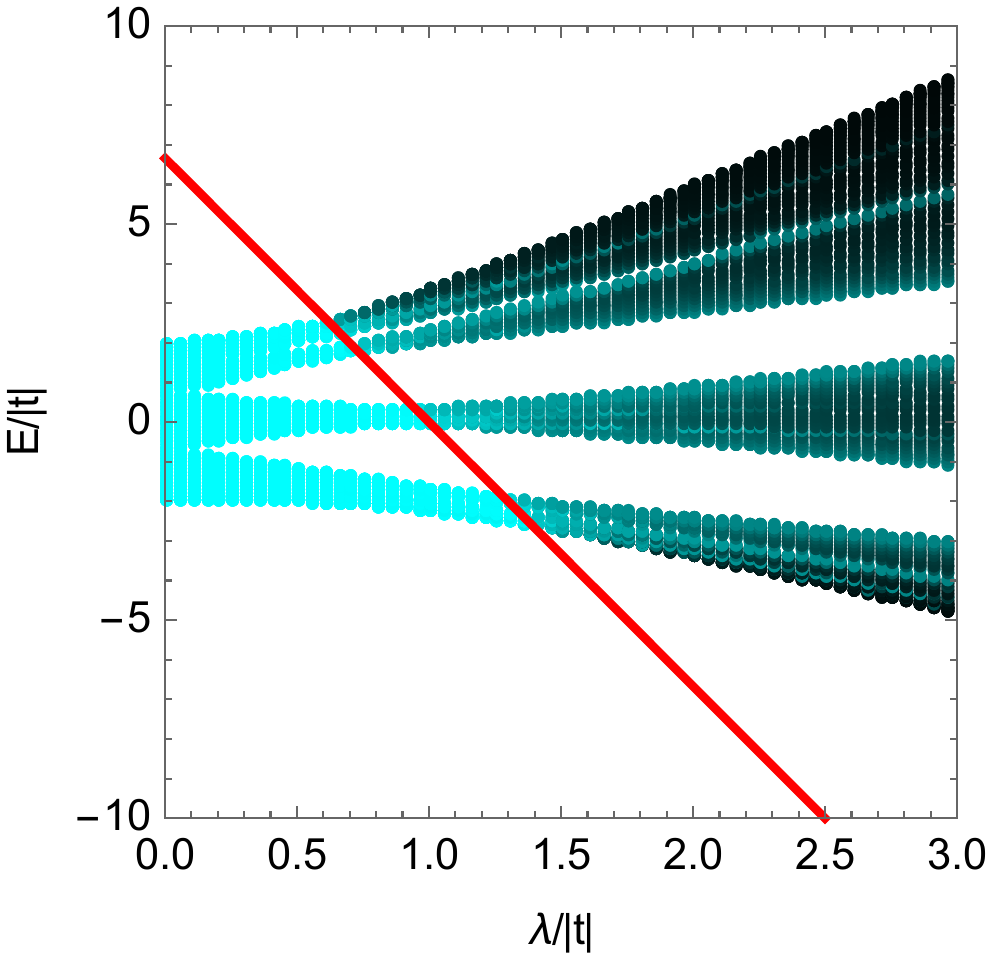}\\
\includegraphics[scale=0.5]{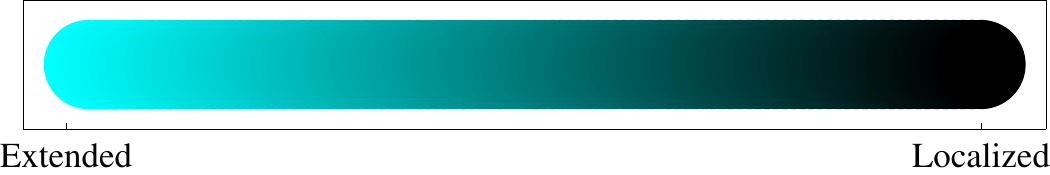}
\caption{\label{gaa}Localization transition in the GAA model. The red line is the self-dual mobility edge point ($\alpha E=2(|\lambda|-|t|)$) at which the energy dependent localization transition takes place. The spectrum is plotted as a function of $\lambda$ for $\alpha=0.3, t=1$. Black stands for $\text{IPR}= 1$,  corresponding to the fully localized state. Cyan denotes the value $1/L$ (where $L=200$ is the number of sites), corresponding to the fully extended state.}
\end{figure}
To establish the self duality in the GAA model we rewrite Eq.(\ref{modelI}) as,
 \begin{align}
t(\psi_{p-1}+\psi_{p+1})+g \chi_p(\beta) \psi_p=(E+2\lambda \cosh \beta)\psi_p,
\label{model1}
\end{align} 
where we have defined the on-site potential $\chi_p(\beta)$ as 
\begin{align}
\chi_p(\beta)&=\frac{ \sinh \beta }{\cosh\beta-\cos (2\pi p b+\phi)}=\sum^{\infty}_{r=-\infty} e^{-\beta |r|} e^{i r( 2\pi p b+\phi)}\\
g &=2 \lambda\cosh \beta/\tanh \beta
\la{id},
\end{align}
with $1/\alpha=\cosh \beta$ for $\alpha>0$, $t>0$, and $\lambda>0$. 
Now we define a generalized duality transformation, 
\begin{align}
f_k=\sum_{m n p}e^{i 2\pi b(k m+m n +n p) }\chi_n^{-1}(\beta_0) \psi_p \la{trans},
\end{align}
with $\cosh \beta_0=\frac{E+2\lambda \cosh \beta}{2t}$. In the $f$ space, the model can be expressed as,
   \begin{align}
t(f_{k+1}+f_{k-1})+g\frac{\sinh \beta}{\sinh \beta_0}\chi_k(\beta_0) f_k=2t\cosh \beta f_k. \la{dualmodel}
\end{align}
The model in the $f$ space is identical to the model in the $\psi$ space for the condition $\beta=\beta_0$. For this condition the duality condition is given as,
\begin{align}
	\cosh \beta=\frac{E}{2t-2\lambda}
\end{align}
This explicit dependence on the energy results in a critical energy as a function of the model parameters that defines the mobility edge for the GAA model. In terms of $\alpha$, the mobility edge separating the localized and extended states for Eq.~(\ref{onsite}) is given by the following extremely simple closed-form expression,
\begin{align}
\alpha E= 2 \sgn(\lambda)(|t|-|\lambda|). \la{mainresult}
\end{align}
Thus the GAA model generalizes the AA duality to include mobility edges in a family of nearest-neighbor quasiperiodic models.

The localization properties of an eigenstate can be numerically quantified using the inverse participation ratio (IPR). The IPR for an eigenstate $E$ is given as,
\begin{align}
\text{IPR}(E)=\frac{\sum_n|\psi_n(E)|^4}{(\sum_n |\psi_n(E)|^2)^2}.
\la{ipr}
\end{align}
For a localized eigenstate,  the IPR approaches the maximum possible value $\sim1$. For an extended state, the IPR is of the order $1/L$, which is vanishingly small in the large system size limit. This behavior across the mobility edge is shown clearly in Fig.~\ref{gaa}. 

\begin{figure*}
\centering
\includegraphics[scale=0.62]{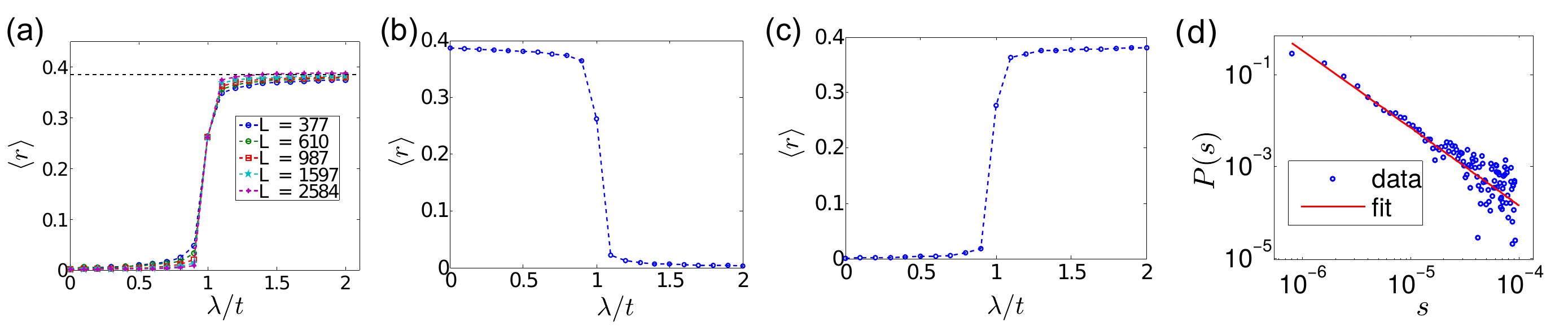}
 \caption{(a) Level statistics of the AA model averaged over eigenstates and $\phi$. The black dashed line is the expected value from Poisson statistics ($\langle r \rangle=2\ln 2-1$). $\langle r\rangle\rightarrow 0$ for large system sizes in the delocalized side ($\lambda/t <1$). 
 (b) Level statistics averaged over the dual flux $\varphi$ (twist angle) with the system size $L = 987$.  (c) $Q$ is set from the Pell sequence and we average over $\phi$ as in (a). (d) shows the probability distribution (with double logarithmic scale) of the level spacing of the AA model at the self-dual point. In (d), we use critical $\lambda/t = 1$, $L = 4181$, $Q/2\pi = 2584/4181$, the shift $\phi$ is averaged over. The distribution $P(s)$ fits well to a power-law function  $\sim s^{-1.68}$. Reproduced from Ref.~\onlinecite{Li_2016_PRB}.
 }
 \label{fig:rvalueAAH}
\end{figure*}

 \subsection{Differences between random and incommensurate single-particle localization}
There are several key distinctions between localization in quasiperiodic models and that of models with random-disorder. The major distinction lies in the nature of the extended states. For a random disorder, the extended states manifest only in 3D (for particular symmetry classes). These eigenstates are diffusive in nature and the corresponding level statistics follows a Wigner-Dyson distribution. The extended states in the AA model are modulated plane waves (ballistic) and the level statistics are dictated by large spectral gaps. On the localized side the level statistics follow a Poisson distribution upon averaging over the phase $\phi$. The level statistics is described by the adjacent gap ratio $r$ defined as
\be 
r_n = \frac{ {\rm min} (s_n, s_{n-1})}{ {\rm max} (s_n, s_{n-1}) }, 
\label{eqn:r}
\ee  
with $s_n=E_n-E_{n-1}$ being the level spacing between the $n$-th and $(n-1)$-th eigenstates. 
The average of $r_n$ is 
either over the phase $\phi$ or over twisted boundary conditions (with twist angle $\varphi$). 
In Fig.~\ref{fig:rvalueAAH}, we show the level statistics across the localization transition averaging over $\phi$ (a) and (c) or over the twist angle $\varphi$ (b). In the dual space, the twist acts like a random phase for an incommensurate potential in momentum space and therefore the level statistics are dual as well, i.e. phase (twist) averaging yields Poisson level statistics in the localized (delocalized) phase. The distribution of level spacings $P(s)$ at the self-dual critical point shows a power-law decay.

The critical properties of the localization transition in random and incommensurate models are rather different. In both case the localization length diverges at the transition following $\xi\sim \delta^{-\nu}$, where $\delta$ is the distance to the critical point and $\nu$ is the localization length exponent. In disorder driven localization, expectation values of observables follow a well defined probability distribution. Following Chayes-Chayes-Fisher-Spencer (CCFS) this leads to a rigorous bound on the correlation length exponent~\cite{Chayes_1986_PRL}, namely for the disorder driven transition to be stable $\nu$ must satisfy $\nu \ge 2/d$. However, in incommensurate models the CCFS criteria does not apply as there is no distribution governing expectation vales and $\phi$ (or $\varphi$) averaging is not necessary. This is rather clear for the AA model, since $\nu=1$ exactly (see Eq.~(\ref{eqn:xi})). It is also a natural question whether the GAA model is in the same universality class as the AA model. This has been computed in Ref.~\onlinecite{Li_2016_PRB} from finite scaling of the IPR
\begin{eqnarray}
\text{IPR}(\epsilon) &\sim& \frac{1}{L^{d_2(\epsilon)}}f(L^{1/\nu} \delta)
\end{eqnarray}
 where $\delta$ is the distance to the critical point, $d_2(\epsilon)$ is the fractal dimension of the wavefunction (that is energy dependent in the AA model~\cite{1992_Hiramoto_AAScaling}, while $\nu$ is universal for all eigenstates), and $f(x)$ is a scaling function. To accurately compute $\nu$ and $d_2$ we have to both get sufficiently close to the critical point either in coupling or energy and reach sufficiently large system sizes. 
 \begin{figure}[h!]
\centering
\begin{minipage}{.4\textwidth}
  \centering
  \includegraphics[width=0.75\linewidth,angle=-90]{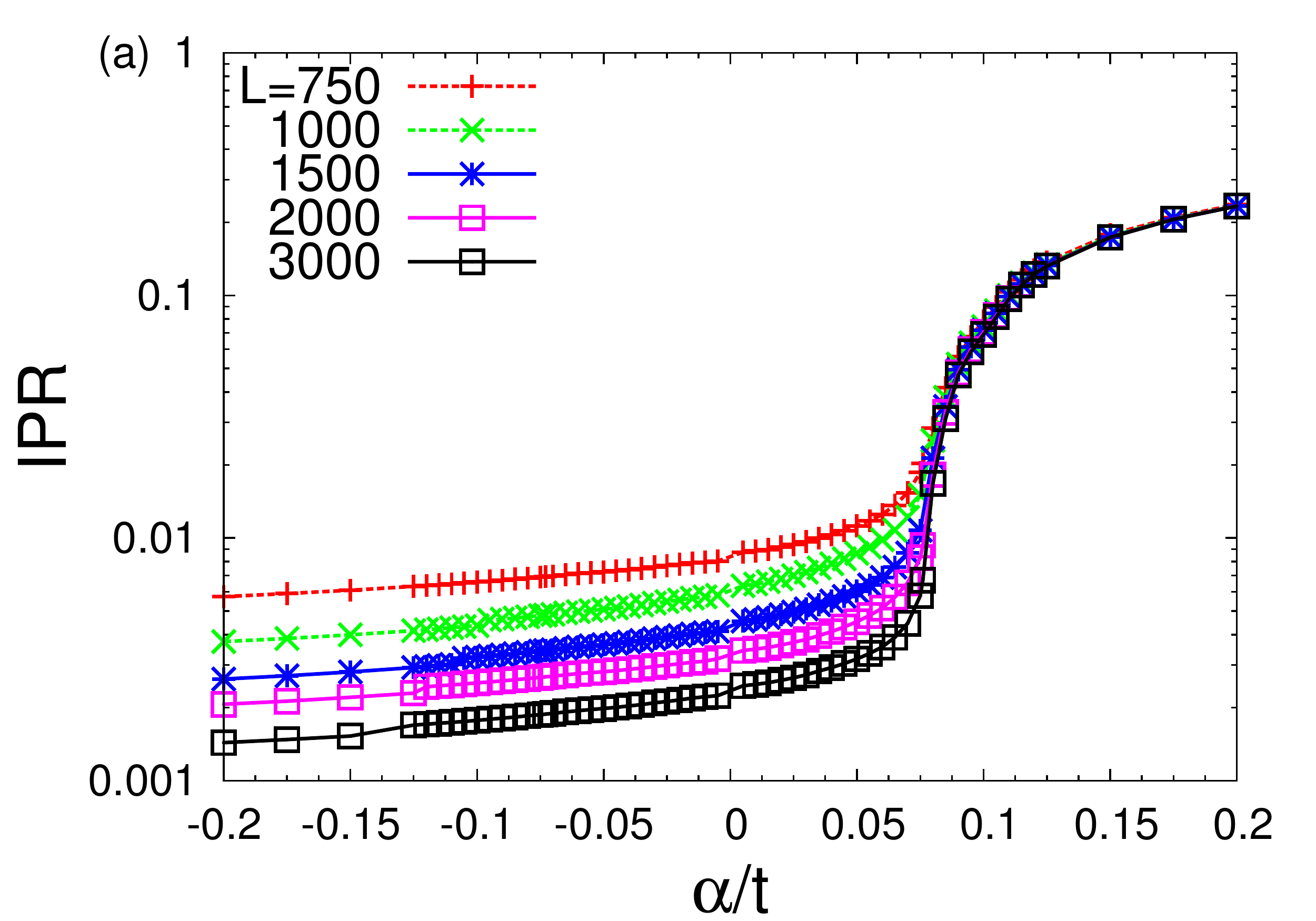}
\end{minipage}
\\
\begin{minipage}{.4\textwidth}
  \centering
  \includegraphics[width=0.75\linewidth,angle=-90]{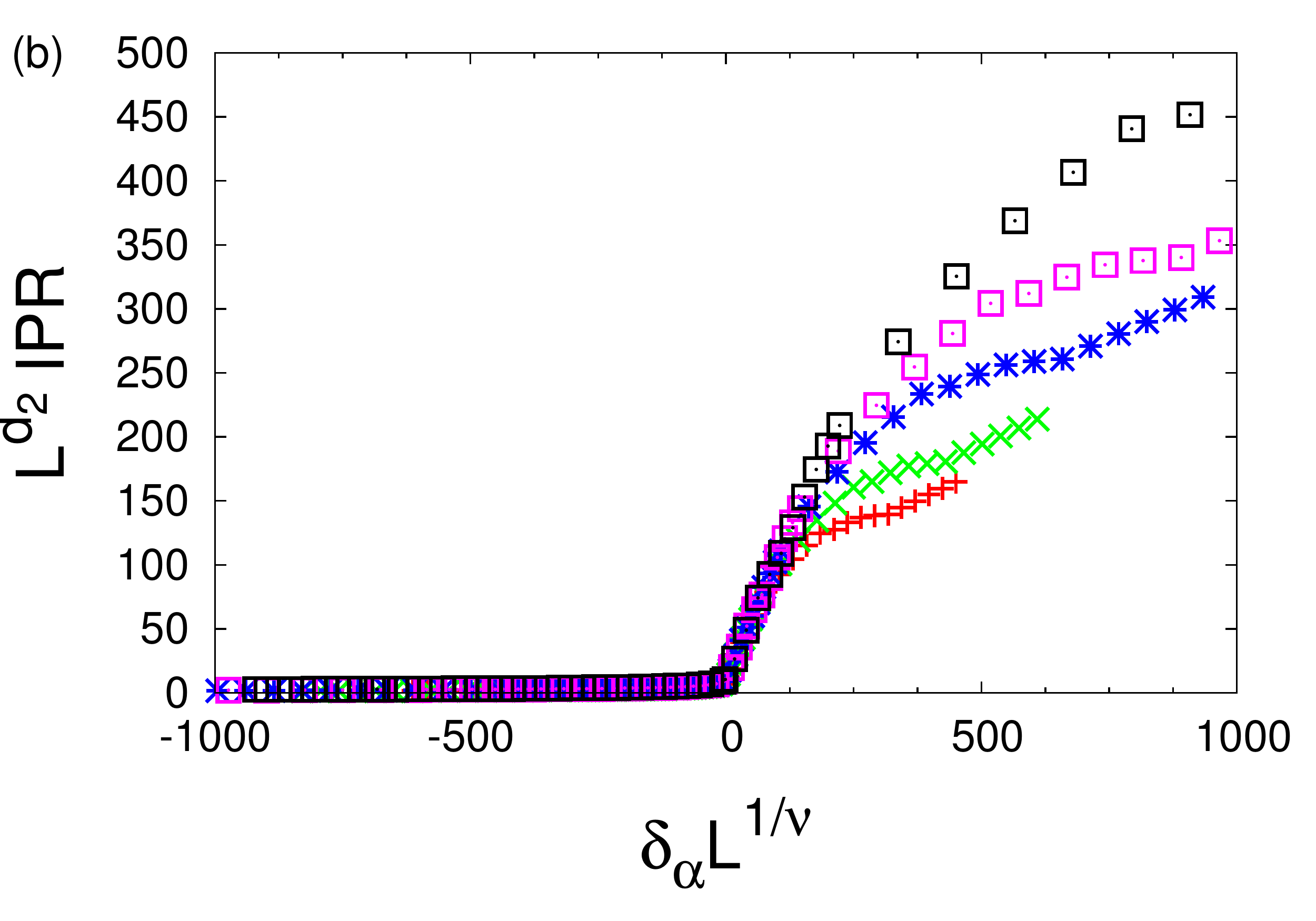}
\end{minipage}
\caption{The IPR near the localization transition in the GAA model for $\lambda=0.9t$ as a function of $\alpha$ and averaging over energies above the mobility edge $\epsilon>\epsilon_{ME}$ . The values of the critical exponents are given in table~\ref{tab:exponents}. Reproduced from Ref.~\onlinecite{Li_2016_PRB}.}
\label{fig:GAA_IPR}
\end{figure}
 However, for the former, the distance to the mobility edge in energy is limited by what energy the eigenstates exist at, and you can have states that do not get sufficiently close to the mobility edge to get good power law scaling. To circumvent this we average over all energies that are on a particular side of the mobility edge, this ``smears out'' our estimate of $d_2$ but produces a very accurate estimate of $\nu$. Following this we find excellent single parameter scaling as shown in Fig.~\ref{fig:GAA_IPR} and the conclusions of this analysis are given in Table~\ref{tab:exponents}.

\begin{table}[]
\begin{tabular}{| c | c | c | }
\hline  & $\nu$ & $d_2$ \\ \hline 
$\alpha = 0$  & $0.98\pm0.05$ & $0.51\pm0.03$  \\ \hline
$\lambda=\lambda_c$  & $0.97\pm0.07$ & $0.51\pm0.03$ \\ \hline
$\lambda = 0.9$ ($E >E_{ME}$)  & $0.95\pm0.08$ & $0.92\pm0.05$ \\ \hline
$\lambda = 0.9$ ($E <E_{ME}$)  & $1.05\pm0.08$ & $0.90\pm0.06$     \\ \hline 
\end{tabular}
\caption{Critical exponents of the generalized AA model taken from Ref.~\onlinecite{Li_2016_PRB}.}
\label{tab:exponents}
\end{table}

 \subsection{Local integrals of motion for quasiperiodic models}

 Localization of eigenstates can also be associated with the existence of {\it local} conserved quantity. This picture connects localization phenomena to the notion of {\it emergent integrability}. This point of view has been particularly useful in demonstrating the existence of MBL as an emergent integrable phase that violates ETH. However, an open question that remains is if a similar construction can be carried out when some single-particle states are extended. Identifying an extensive number of conserved quantities in the presence of thermalizing channels would point to a more robust violation of ETH. In the following we construct the LIOM for a quasiperiodic model in presence of a mobility edge. For the GAA model with the mobility edge we only have partial set of LIOM associated with the localized states. In the following we construct explicit local conserved charges for the GAA model with a mobility edge. For convenience, we rewrite Eq. (\ref{modelI}) as
\begin{align}
H_{GAA} &=\sum_i\mu_ic^{\dagger}_ic_i-y\sum_{ij}t c^{\dagger}_ic_{i+1}+h.c,	\\
\mu_i &=2\lambda\frac{\cos(2\pi i b)}{1-\alpha \cos(2\pi i b)}.
\label{model}
\end{align}
We then construct conserved charges $Q(l)$ systematically in powers of $y$. The convergence of the power series in $y$ determines the existence of local conserved charges corresponding to the localized states. Consider the following form for $Q_0(l)$,
\begin{align}
Q_0(l)=P_0(l)+y P_1(l)+y^2 P_2(l)+.....,\\
P_0=n_0, \, \ \ P_m=\sum_{ij}\eta^m_{ij}(l)(c^{\dagger}_ic_j+c^{\dagger}_jc_i).	
\label{q0}
\end{align}
\begin{figure}[htp] 
\includegraphics[scale=0.5]{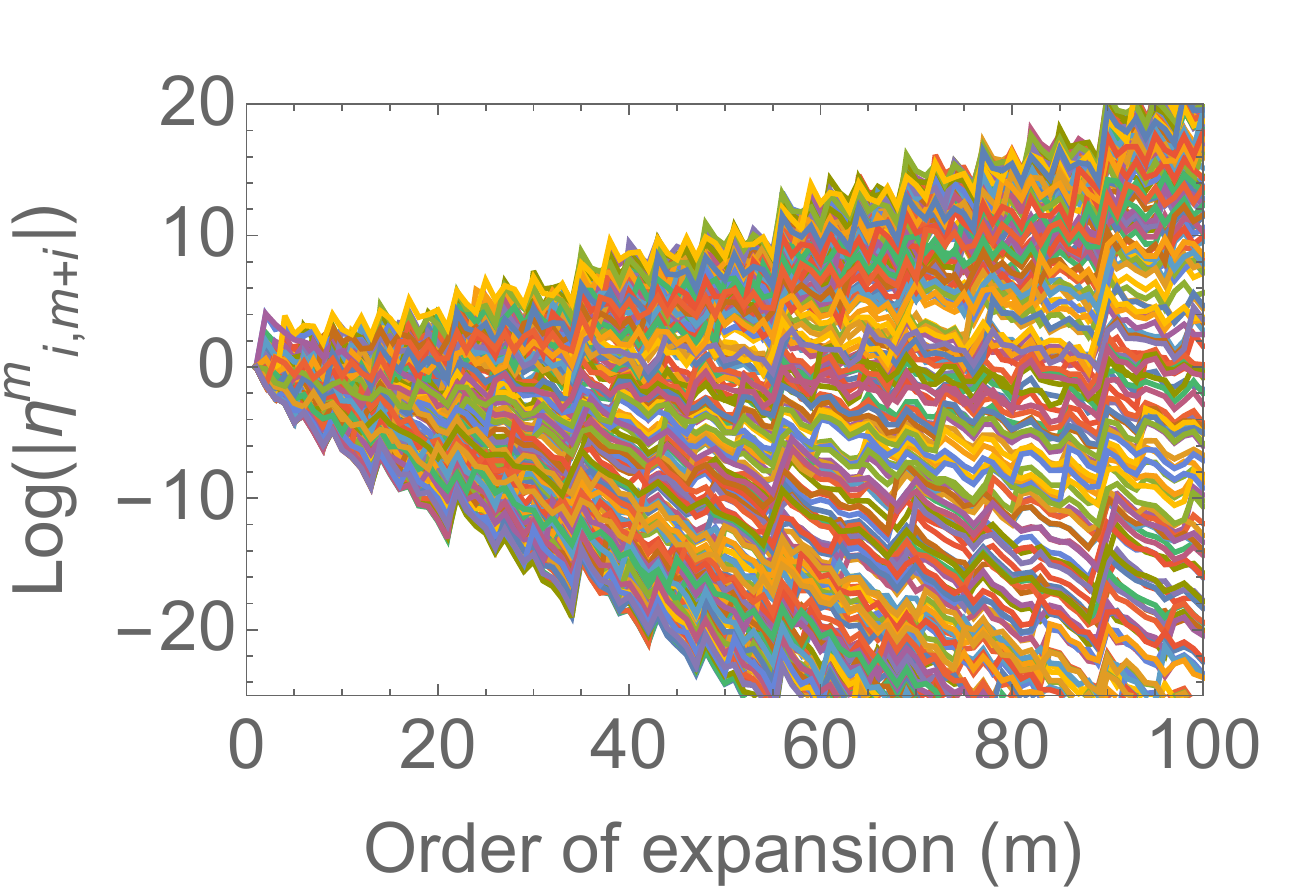}
\caption{Convergence of LIOM: Plots of the typical term  $\log|\eta^m_{l,m+l}|$ ($l=0,...,199$) for $\lambda/|t|=1.0$.}
\label{fig:liom1} 
\end{figure}
This method was first developed in the context of constructing LIOM for the MBL Phase~\cite{Ros_2015_NPB}. Recently this method was applied to quantify the localization transition in the AA model in Ref.~\onlinecite{modak2015integrals} and was subsequently applied to quantify mobility edge in the GAA model in Ref.~\onlinecite{Li_2016_PRB}. 
$P_m$ is the most general quadratic operator that commutes with the  Hamiltonian. The coefficients $\eta^m_{ij}(l)$ are self-consistently evaluated such that it enforces commutation of the conserved charges with the Hamiltonian and among itself. To detect the mobility edge, we need to compute all charges $Q_0(l=0..N-1)$,
\begin{align}
[Q_0(l), H]=[P_0(l), H]+&\sum^{\infty}_{m=0}y^{m+1}\\
\times &([P_m(l), H_1]+[P_{m+1}(l),H_0]),\nonumber\\
\eta^{m}_{ij}=\frac{t}{\mu_j-\mu_i}(\eta^{m-1}_{i,j-1}+\eta^{m-1}_{i,j+1}&-\eta^{m-1}_{j,i-1}-\eta^{m-1}_{j,i+1}), \ \forall (i<j).
\end{align}
The above recursion in $\eta^m(l)$ can be computed order by order for all terms given some initial conditions. The recursion structure is exactly the same for different charges except at the initial condition,
\begin{align}
 \eta^0_{ll}=1,\ \ \  \eta^m_{ij}(l)=\eta^m_{ji}(l), \ \ \eta^0_{ij}(l)=0 \,  (\forall i\ne0,\ j\ne 0).
\end{align}
This recursion generates growing number of hopping terms with increasing order of expansion $m$. The convergence of the power series is indicated by the typical term $\eta^m_{l,m+l}$. For a choice of parameter where the GAA model manifests a mobility edge, only some of the charges would converge corresponding to the localized states. Fig.~\ref{fig:liom1} shows the convergence of the typical terms for $\lambda/|t|=1.0$ and $\alpha=0.3$. The number of converged charges matches the number of localized states computed from the exact expression for the mobility edge~\cite{Li_2016_PRB}.

\subsection{Many-body states of free fermions in the presence of a single-particle mobility edge} 
\begin{figure}
\includegraphics[width=.5\textwidth]{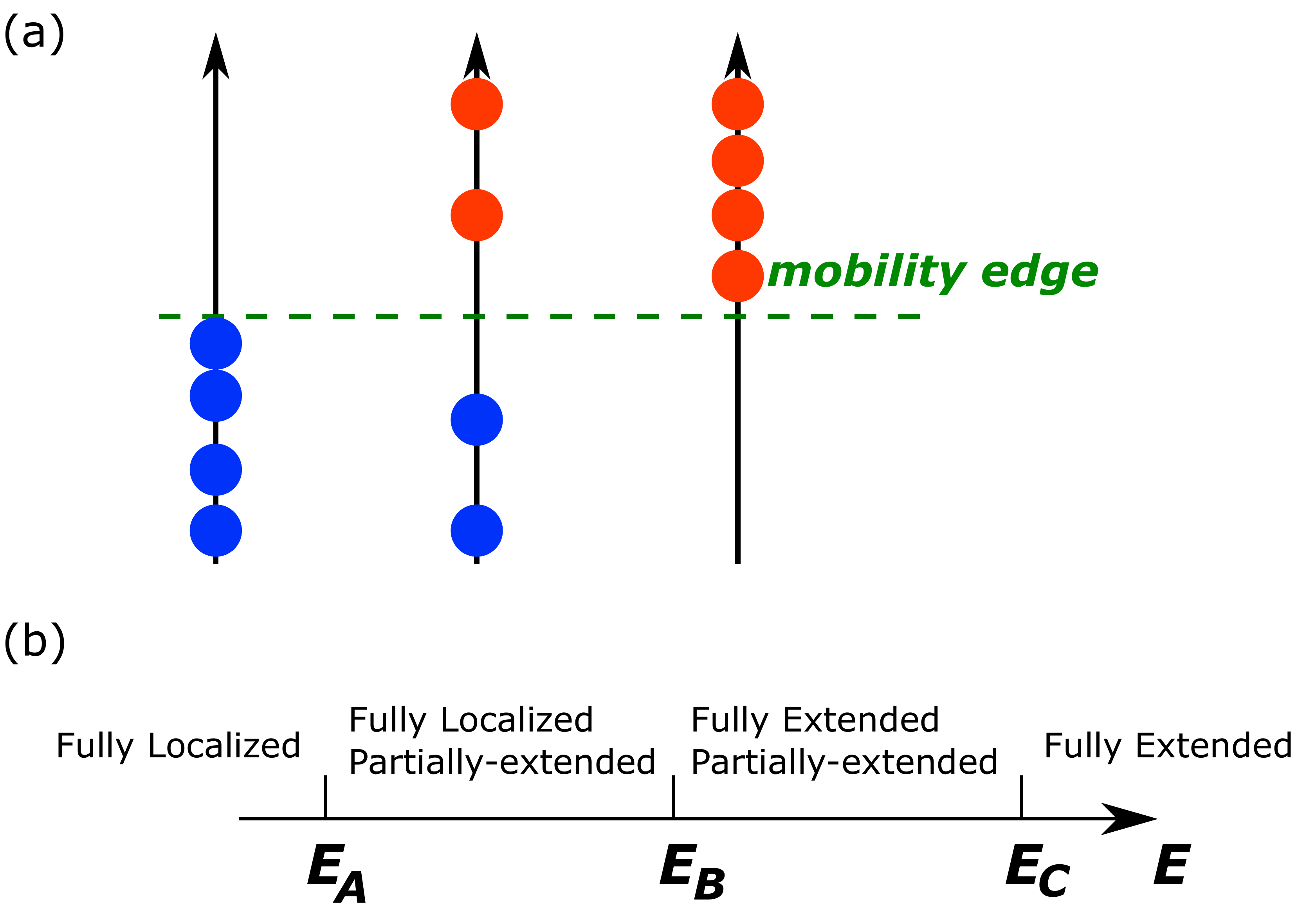}
\caption{Illustration of many-body states of free fermions in the presence of a single-particle mobility edge. (a) shows the three different types of states---fully localized, partially extended, and fully extended. (b) shows different regions of the many-body energy spectra. 
} 
\label{fig:MBfreefermion}
\end{figure}

We now discuss forming many-body eigenstates out of non-interacting single-particle orbitals.
Without interactions, the many-body eigenstate is a product state of single-particle orbitals, 
\be 
|\Psi\rangle _{\rm free} = | m_1, m_2, \ldots, m_N \rangle 
= \psi_{m_1} ^\dag \psi_{m_2} ^\dag \ldots \psi_{m_N} ^\dag | {\rm vac} \rangle,  
\ee 
where $\psi_m$ is the annihilation operator for the single-particle eigenmodes. In the presence of a single-particle mobility edge, there are three different ways of constructing many-body states (Fig.~\ref{fig:MBfreefermion}(a)): (i) all particles put in the localized single-particle orbitals, (ii) all particles put in mobile orbitals, and (iii) some particles put in localized and others in mobile orbitals, which respectively give fully localized, fully extended, and partially-extended many-body states.  
These partially extended many-body states have important consequences. Consider a model with single-particle energies 
$\epsilon_1 < \epsilon_2 < \ldots < \epsilon_{L^d}$ 
with a single-particle mobility edge $\epsilon_{m^\star}$, such that the states with $\epsilon_{m\le m^\star}$ ($\epsilon_{m>m^\star}$) are localized (mobile). Let's consider particle number $N$ satisfies $N\le m^{\star}$ and $N<L^d - m^{\star}$ for simplicity.  We have three relevant many-body energy scales ($E_{A,B,C}$)  separating four different energy windows in the spectra (Fig.~\ref{fig:MBfreefermion}(b)). The first one $E_A$ is the lowest energy of partially-extended states given by 
$$ 
E_A =  \epsilon_{m^\star  +1} + \sum_{p =1} ^{N-1} \epsilon_{p},   
$$ 
below which the eigenstates are all fully localized. 
The second one $E_B$ is the maximal energy 
of the fully localized states given by 
$$ 
E_B = \sum_{p=1}^{N}  \epsilon_{m^\star -p+1},   
$$ 
above which the eigenstates are either fully or partially extended. 
The third energy scale $E_C$ is the the maximal energy of partially extended states, 
$$
E_C  =  \epsilon_{m^\star} +  \sum_{p=1}^{N-1} \epsilon_{L^d - p+ 1},   
$$ 
above which the eigenstates are all fully extended. The resultant physical picture of many-body energy spectra  is shown in Fig.~\ref{fig:MBfreefermion}(b). The energy windows on the sides (below $E_A$ or above $E_C$) are actually intensive, thus tiny in the thermodynamic limit. It is worth noting here that the single-particle mobility edge does not convert to a many-body mobility edge even in absence of interactions, as one may intuitively expect. 
 
The partially extended many-body states have several unique properties, making them different from either fully localized or fully extended ones. 
First, the many-body normalized participation ratio (NPR)~\cite{Iyer_2013_PRB} measures how the many-body wave function ``spreads'' over Fock space, which indicates nominally a non-conducting behavior as shown analytically in Ref.~\onlinecite{Li_2015_PRL}. For an ergodic system eigenstate wave functions are expected to approximately explore the entire Hilbert space, giving rise to  a finite NPR. For a non-ergodic system, the wave functions are not able to spread over the whole Hilbert space, 
for which the NPR vanishes. Taking a fully localized state, the NPR vanishes following a finite-size scaling form 
$ 
\eta \propto 1/ V_H .  
$ 
For a partially-extended state, the NPR vanishes according to a different scaling formula 
\be 
\eta \propto 1/V_H ^{\zeta},  
\ee 
with the exponent $\zeta \in (0, 1)$.  This behavior of partially extended states is similar to the non-ergodic extended phase studied in the context of single-particle localization in  Bethe lattices~\cite{2012_Biroli_arXiv,2014_Luca_BetheLattice_PRL}.  
Second, the partially extended states have extensive but subthermal entanglement entropy. For an ergodic system, the entanglement entropy is expected to reflect the thermal entropy.  For the partially extended states, the entanglement entropy obeys the volume law due to the extended orbitals. At the same time, the entanglement entropy strongly deviates from the thermal entropy because of the localized orbitals~\cite{Li_2015_PRL,Li_2016_PRB}, which indicates  nonthermal and nonergodic character. 
Third, the local observables in partially extended states show ETH violation. For a system respecting ETH, the fluctuations of local observables among nearby eigenstates are suppressed. In the partially extended phase, the local observables exhibit strong fluctuations (with $\sqrt{N}$ scaling), which have been analytically worked out and numerically verified for 1D incommensurate and 3D Anderson lattice models~\cite{Li_2016_PRB}.  The non-thermal fluctuations originate from localized orbitals composing the partially extended many-body states.

From the LIOM perspective, the many-body model of free fermions with a single-particle mobility edge has an subextensive number of LIOM (their number is proportional but smaller than the system size). By using these LIOM, one cannot completely diagonalize the many-body Hamiltonian.  The Hamiltonian can actually be brought to a block-diagonal matrix through local-unitary transformations, with each block of dimension $2^{\alpha L^d}$ ($\alpha <1$). The presence of extensive number of LIOM gives rise to nonergodic behavior and the exponential dimension of the Hamiltonian-block indicates the system is extended in real-space. We thus have a non-ergodic metal. Whether this subextensive LIOM pictures survives against interactions is still an open question.  The essential challenge is to bound the delocalization instabilities from resonances in presence of extended degrees of freedom, which might be more difficult than proving the existence of MBL~\cite{Imbrie_2016_JSP}

\section{ Many-body localization with a single-particle mobility edge}
The interacting AA model has been studied both numerically~\cite{Iyer_2013_PRB} and analytically~\cite{Mastropietro_2015_PRL}.  The persistence of ground state localization in the presence of a weak many-body interaction has been established in a mathematically rigorous way ~\cite{Mastropietro_2015_PRL} and numerical results have provided strong evidence that the localization is stable against interactions for the whole spectra. Numerically, interactions are found to make the critical incommensurate potential larger, meaning interactions make the system more difficult to localize. It is however worth mentioning that the change in the phase boundary due to interaction effects is rather small in the numerics~\cite{Iyer_2013_PRB}.  In contrast, the GAA model does have a single-particle mobility edge (see Sec. II)
which makes the effects of  interactions more difficult to analyze, but at the same time, makes the physics richer. The interplay 
of coexisting localized and extended orbitals in interacting systems is particularly interesting.

We now review some recent developments in this area. First, we discuss the absence of MBL 
for disordered systems with 
intrinsic or symmetry-protected topological delocalized states, such as disordered integer quantum Hall insulators~\cite{nandkishore2014marginal}. 
We then discuss the scenario where the delocalized states are not protected such as systems with single-particle 
mobility edges. We review some recent numerical studies~\cite{modak2015many,li2015many,modak2016criterion} which suggest that a non-ergodic phase 
can exist in 1D models with 
quasiperiodic potentials which have single-particle mobility edges. Next, we analyze the possibility of this non-ergodic phase being delocalized (i.e. a non-ergodic metal)\cite{li2015many,Li_2016_PRB} in these models. Finally, we review 
MBL  in the presence of an ergodic bath. Usually a bath is taken to be very large and the back action of the system which is in contact with
it is negligible.  Some recent studies have found that if the system is very strongly localized and the bath is 
weakly ergodic, the system can even localize the bath, 
through the MBL proximity effect~\cite{nandkishore2015many,hyatt2016many}.

\subsection{Thermalization in the presence of protected delocalized states } 

\begin{figure}
\includegraphics[width=3.2in]{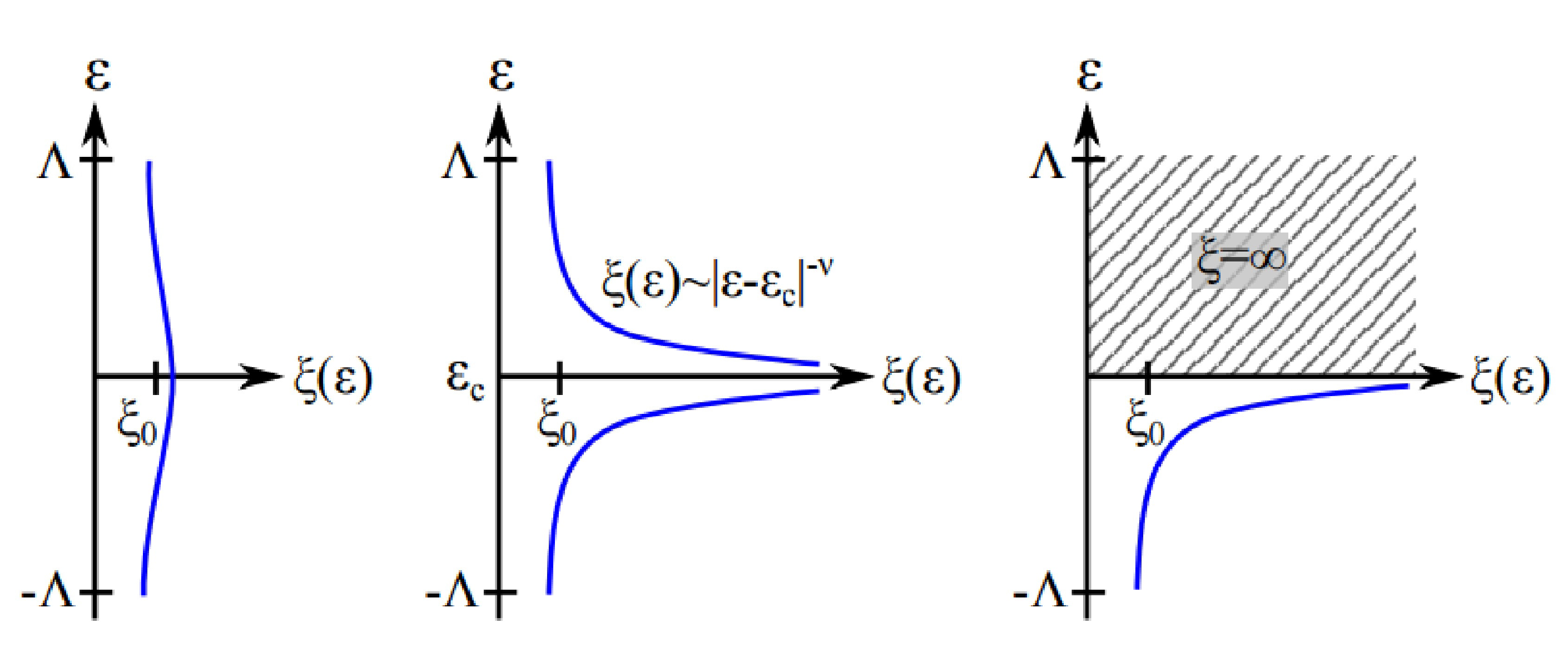} 
\caption{Schematic of dependence of localization length for (A) completely 
localized system, (B) marginal system and (C) system with mobility edge.(Figure 
taken from ~\cite{nandkishore2014marginal})}
\label{Fig: ll}
\end{figure}

Nandkishore and Potter have addressed the issue of whether topological edge states can be protected 
by the localization of the bulk states~\cite{nandkishore2014marginal}. Specifically, they studied marginally localized systems with only one critical energy eigenstate 
$E_c$ and single-particle eigenstates with energy $E$ and diverging localization length $\xi(E)\sim\frac{1}{|E-E_c|^{\nu}}$ near $E_c$ (see Fig.~\ref{Fig: ll}). 
Disordered integer quantum Hall systems~\cite{thouless1984wannier,huckestein1995scaling}, chiral superconductors, intrinsically topological
superconductors~\cite{motrunich2001griffiths}, and  topological insulators with symmetry preserving
disorder~\cite{soluyanov2011wannier} are examples of such marginally localized systems. We discuss below some of the conclusions of this study and related ones.  

\subsubsection{Non-interacting marginally localized systems}
Non-interacting marginally localized systems are unique in themselves. They show anomalous sub-diffusive dynamics~\cite{sinova2000liouvillian} 
with quantum  Hall  systems being an example~\cite{moore2003conservation}.
The sub-diffusive dynamics  can be thought of in terms of a length scale dependent diffusion 
constant, $D(L)\sim L^{-2/(2\nu-1)}$, which leads to a length scale dependent DC conductivity,
$\sigma(L)\sim L^{-2/(2\nu-1)}$. Hence, in the thermodynamic limit $\sigma(L)$ vanishes 
 algebraically for such marginal systems, which is interestingly an intermediate scaling between $\sigma (L)\sim \mathrm{const}$ 
(for systems with extended states and diffusion) and $\sigma (L)\sim \exp(-L/\xi)$ (for systems with localized states and no diffusion).

The entanglement scaling of a marginal system is also quite different when comparing completely (i.e. the entire spectrum) extended and localized systems.
It is well-known that for completely localized systems, the entanglement entropy of an  
excited energy eigenstate obeys an area law~\cite{eisert2010colloquium} $S(R)\sim \xi R^{d-1}$, whereas, 
for completely extended  systems it follows a volume law~\cite{amico2008entanglement} $S(R)\sim  R^{d}$,
where $R$ is the linear size of a subsystem  and $\xi$ is the  localization  length associated with the state.
A marginal  system with diverging localization length exponent $\nu$ can be thought of as a system where 
a fraction $f(R)\sim R^{-1/\nu}$ states are extended across a subsystem of size $R$ which makes 
a volume-law contribution to the entanglement entropy and the rest of the states make an area law contribution. Hence, $S(R)\sim f(R)R^{d}+(1-f(R))\xi R^{d-1}$.
For $\nu>1$, $f(R)R^{d}$ dominates over the area law and $S(R)\sim R^{d-1/\nu}$ obeys
a fractal scaling~\cite{nandkishore2014marginal}. However, for $\nu <1$ the entanglement entropy should obey an area law.

\subsubsection{The effect of interactions on marginal systems}
 
The eigenstates of a many-body localized system have a LIOM description~\cite{serbyn2013local,huse2014phenomenology}. 
The LIOM for non-interacting systems can be expressed simply in terms of 
single-particle orbitals. In an MBL
phase, these LIOM are dressed by interactions. 
Using the same analogy, one can argue that marginally localized systems have  a
(nearly) complete set of integrals
of motion which are algebraically localized. This means
that a fraction of the integrals of motion decay as a power
law in $R$, in contrast to traditional MBL systems (where integrals of motion decays
exponentially in $R$) and  a delocalized
system (where the integrals of motion do not converge).

Nandkishore and Potter have investigated the fate of such marginally localized 
systems after the introduction of interactions.
The system is described by the Hamiltonian, $H=H_0^M+V$,
where, $H_0^M$ corresponds to the Hamiltonian of the non-interacting marginal system
and the interaction term $V$ can be described by,$V=\sum_{\alpha \beta \gamma \delta}
\lambda_{\alpha \beta \gamma \delta}\psi_{\alpha}^{\dag} \psi_{\beta}^{\dag} \psi_{\gamma}\psi_{\delta}$,
where, $\psi_{\alpha}^{\dag}$ ( $\psi_{\alpha}$) creates (annihilates) a single-particle eigenmode of 
energy $E_{\alpha}$.
$|\phi(E) \rangle$ is used to denote a many-body eigenstate of $H_0^M$ 
of energy $E$ labeled by a particular set of marginally localized
integrals of motion. $|\psi(E) \rangle$ is used to express a many-body eigenstate of energy $E$  of
an interacting Hamiltonian $H=H_0^M+V$ . The operator $T$ is defined as,  $V|\psi(E) \rangle=T|\phi(E) \rangle$. 
$T$ can be expressed in terms of $V$ using the Dyson equation~\cite{mahan2013many},
\begin{equation}
T=V+V\frac{1}{E-H_0^M+i0}T.
\end{equation}
Using perturbation theory when $V$ is small, $T$ can be calculated using the Born series,
$T=V+VG_0V+VG_0VG_0V+...$, with $G_0=1/(E-H_0^M+i0)$.
Matrix elements of $T$ between single-particle states with energy
$E$ and spatial separation $R$ give the effective matrix elements for long 
range hopping.  If the effective hopping matrix
elements  fall off with distance faster than $1/R^{d}$,
localization is  possible~\cite{anderson1958absence}.  Similarly, the matrix elements of
$T$ between two or higher particle states with the
particles  separated  by  a distance $R$ generate again the effective
matrix  element  for long range two-body or higher body interactions.
Again, there exist conditions on how fast these matrix elements must fall off with
$R$ such that  localization becomes stable~\cite{fleishman1980interactions,levitov1990delocalization}.

In other words, if $|\Psi_1\rangle$ is  an  eigenstate  of $H_0$ with  localization length $\xi_1$ and the probability that
$|\Psi_1\rangle$ is in resonance with another eigenstate $|\Psi_2\rangle$ a distance $R$ apart 
($R>>\max(\xi_1,\xi_2)$) goes  to  zero as $R\to \infty$, then algebraically
localized integrals of motion can be defined to arbitrary precision.
In contrast,  if the probability of having a resonance
at a length scale greater than $R$
does not go to zero as $R\to \infty$, then algebraically
localized integrals of motion do not exist, and
the system cannot support marginal MBL.
Nandkishore et al. have investigated how  the matrix  elements of $T(R)$ scale with distance $R$ for different type of resonance processes
case by case in their work.  The first order term $O(V)$ is purely short-ranged. The first nontrivial  long-range term appears at  order $O(V^2)$.

\subsubsection{Flip-flop assisted hopping}
\begin{figure}
\includegraphics[width=3.2in]{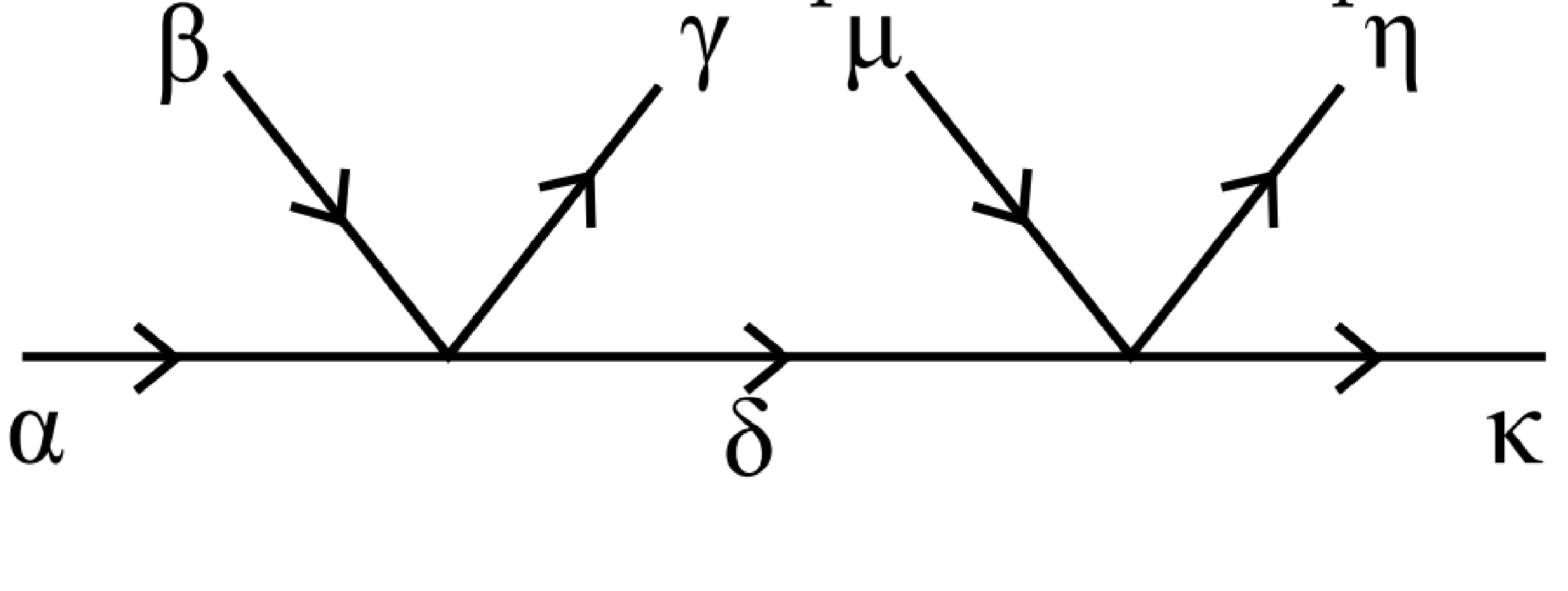} 
\caption{Diagram for flip flop assisted hopping process. (Figure taken from ~\cite{nandkishore2014marginal})}
\label{Fig: flipflop}
\end{figure}

The matrix elements of $T$ between localized single-particle orbitals (hopping resonances)
and the matrix elements of $T$ between two particle states (flip-flop resonances) 
fall off sufficiently rapidly with $R$  such that the
probabilities of having long-range hopping resonances and flip-flop resonances vanishs for any finite $\nu$. 
Thus, these processes do not present an obstacle to the construction of marginally localized integrals of motion.

Flip-flop assisted hopping processes are those which describe 
the hopping of a particle between two points separated by a distance $R$ by
triggering single-particle transitions at both points (see Fig.~\ref{Fig: flipflop}).   The matrix element of 
a flip-flop process is given 
by 
\begin{equation}
T _{\alpha \beta \mu,\gamma \eta \kappa} (R)= \sum_{\delta} \frac{\lambda_{\alpha \beta \gamma \delta} \lambda_{\delta \mu \eta \kappa}}{E_\alpha + E_\beta - E_\gamma - E_\delta} \nonumber \\
\sim R^{-\frac{1}{2}(d+1/\nu)},
\end{equation}
where the indices label states of the non-interacting model that are localized at different points in the system and $E_\alpha$ is the energy of a state labeled 
by $\alpha$. The $\lambda$'s 
are the matrix elements of the interaction between the localized states and the $({\alpha,\beta, \mu})$ and 
$({\gamma,\eta, \kappa})$ are two triads of 
proximate localized states separated by a distance $R$.  
Localization is expected to be destroyed when $T(R)$ falls off more slowly than $1/R^{d}$.
Hence, when $\nu d >1$ localization will definitely be destroyed. The CCFS criterion~\cite{Chayes_1986_PRL} 
suggests that  $\nu\ge 2/d$ for a system with uncorrelated disorder, which ensures the fact that in the case of a generic marginal
system, the localization is unstable.

\begin{figure}
\includegraphics[width=2.3in,angle=-90]{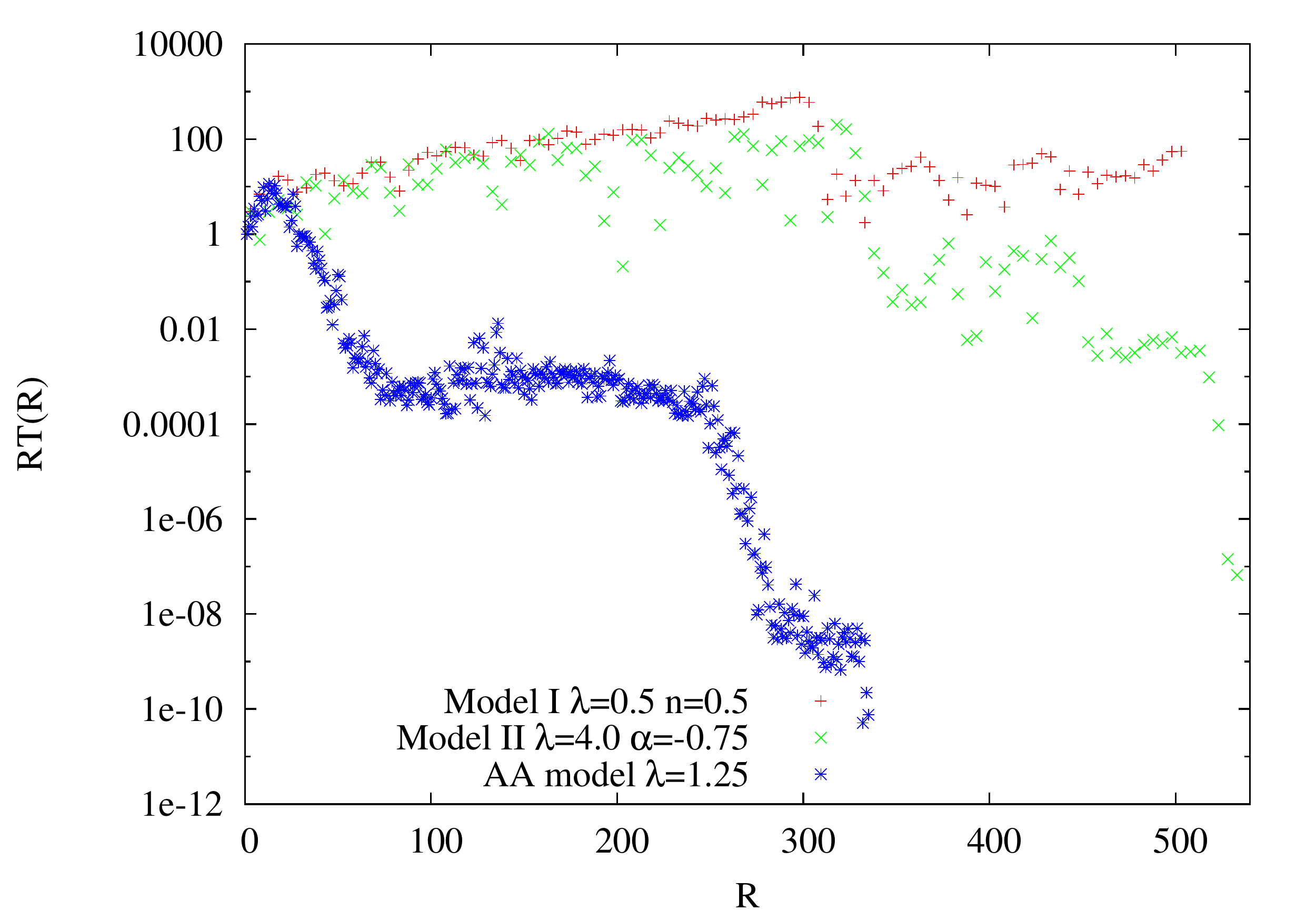}
\caption{Variation of 
$RT(R)$ as a function of $R$ for different models. }
\label{Fig: matrix elements}
\end{figure}

 In order  to investigate the behavior of $T(R)$
as a function of $R$ for microscopic models with single-particle mobility edge,  we consider  models of the form
\begin{equation}
H_0 = - t\sum_j \left( c_j^\dagger c_{j+1} + {\rm h.c.} \right) + \sum_j \mu_j c_j^\dagger c_j,
\label{Eqn: mainham}
\end{equation}
where, $c_i$ ($c_i^\dagger$) annihilates (creates) a particle at site $i$ (which can be thought of as spinless fermions for 
concreteness even though their statistics are irrelevant as far 
as the single-particle spectrum is concerned). $t$ is the nearest-neighbor hopping amplitude and $\mu_j$, an on-site potential. We consider two different types of potentials defined by 
\[  \mu_j = \left\{ \begin{array}{ll}
              2 \lambda \cos \left(2\pi b j^n \right), & \mbox{Model I};\\
              2 \lambda \frac{\cos \left(2\pi b j \right)}{1-\alpha \cos \left(2\pi b j \right)}, & \mbox{Model II}. \end{array} \right. \]
where $b$ is a an irrational number, $0 < n \leq 1$, $|\alpha| < 1$. 
Note that model II is the GAA model already introduced in Eq.~(\ref{model}).
Both models produce quasiperiodic potentials and reduce 
to the AA model in the appropriate limits, $n=1$ for model I and $\alpha=0$ for 
model II (Refs.~\onlinecite{dassarma.1990,ganeshan.2014}). 
In addition, both of these models posses mobility edges.
All single-particle states of model I, with energy between $\pm 2|t-\lambda|$ are delocalized and 
all other states are localized. In the case of $\lambda>t$ all single-particle states are localized as in the usual AA model.
For model II (i.e. the GAA model), there is a mobility edge separating, localized and extended states at an energy $E$ given by Eq.~(\ref{mainresult}), which we also repeat here for clarity,
$\alpha E=2(|t|-|\lambda|)$.

Numerical calculations of  $T(R)$ for models I, II, and AA are shown in Fig.~\ref{Fig: matrix elements}.
It can be seen that while $T(R)$ appears to be falling off as $1/R$ or more slowly for model I, 
the fall off is definitely faster than $1/R$ for model II. $T(R)$ for 
the AA model~\cite{aubry.1980}, which exhibits MBL~\cite{huse.2013} 
in the presence of interactions, falls off much faster than $1/R$.
However, note that since  delocalized states of these models 
are not protected,  arguments of many-body delocalization by Nandkishore and Potter cannot be directly applied  
to these models. Moreover, as we have already discussed the CCFS criteria is not valid for  models with
quasi-periodic potentials.

\subsection{Microscopic models with a single-particle mobility edge} 
The above discussion concerns the fate of localization when interactions couple localized states to a single (or band of) 
delocalized states. The delocalized states can be  considered to be a thermalizing bath for the localized states and whether MBL persists or not 
is analyzed within a framework where the  bath was taken to be very large and the back action of the localized system on it is neglected~\cite{nandkishore2014spectral,johri2015many}. 
The delocalized states are thus by construction protected against localization themselves and it can be argued that 
 a thermodynamically large bath will cause the full system to thermalize even in the presence of an arbitrary weak coupling 
 between the bath and the localized system. Thus, there is no MBL in such systems.
 
 This raises the important question of what happens in a system with localized and delocalized single-particle states upon the 
 introduction of interactions when the effect of each band on the other has to be considered. In other words, what is the 
 effect of interactions in a system when there is no protection for either the localized or delocalized states? One ideal setting to address this question is provided by introducing interactions to models I and II. Focusing on spineless fermions, this leads to the many-body Hamiltonian
 \begin{equation}
 H = H_0 +V\sum _i n_in_{i+1},
 \end{equation} 
which has been solved using exact diagonalization in Refs.~\onlinecite{modak2015many, li2015many} using
 several different MBL diagnostics. We summarize below the conclusions drawn from some of these studies.

\begin{figure}
\includegraphics[height=3.2 in,width=2.7in,angle=-90]{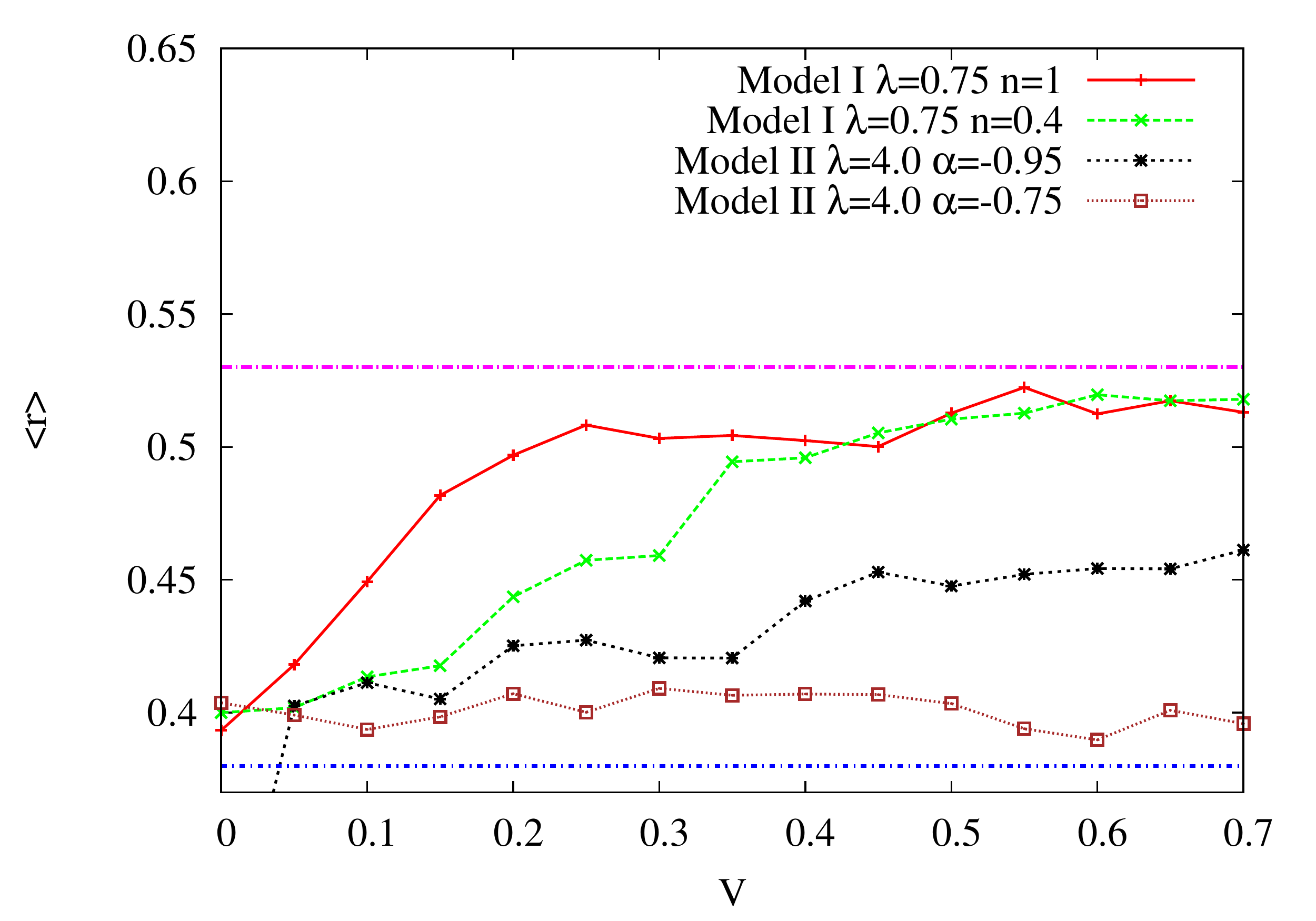}
\caption{The variation of the mean of the ratio between adjacent gaps in the spectrum for $L=14$ at half filling
for model I and model II. The blue dotted line is for the Poissonian distribution and the pink one is
for the Wigner-Dyson distribution. Figure is taken from Ref.~\onlinecite{modak2015many}.}
\label{Fig:level spacing}
\end{figure}

The absence of level repulsion has been extensively exploited as a diagnostic of the MBL 
phase~\cite{pal.2010,huse.2013}. 
In the presence of a finite interaction strength $(V \neq 0)$, the average adjacent gap ratio [defined in Eq.~(\ref{eqn:r})] of model I 
obeys the Wigner-Dyson distribution but for model II it is Poissonian, a signature of the MBL phase (see  Fig.~\ref{Fig:level spacing}). Level 
statistics thus seem to suggest that model I is ergodic while model II remains localized in the presence of interactions.

\subsubsection{Time evolution of the energy averaged entanglement entropy}
The growth of the entanglement starting from an equally weighted sampling of all possible initial product states 
has been argued to be linear in time  in the ergodic phase and much more slower (logarithmic)
in the many-body localized
phase~\cite{bardarson.2012,serbyn.2013}. 
 Assuming a system of length $L$ devided in two equal parts $A$ and $B$, 
the second-order R\'{e}nyi entropy $S_2(t)$ (Ref.~\onlinecite{rrnyi1961measures}) defined as  
$S_2(t)=-\log_2(\text{Tr}_{A}{{\rho_{A}(t)}^{2}})$ with $\rho_A(t)$ 
the reduced density matrix of subsystem $A$ at time $t$, is one of the most extensively used diagnostics to characterize 
the entanglement between two subsystems. 
It is known that in the ergodic phase, $S_{2}(t)\sim t$ at long 
times and saturates to the infinite temperature thermal value $S_2\sim L/2$ (Ref.~\onlinecite{huse.2013}). For 
the  MBL phase , $S_2(t)\sim \zeta\log(t)$, where $\zeta$ is the localization length 
and the saturation value is much smaller than the thermal value ($L/2$), but is still
extensive in system size.

\begin{figure}
\includegraphics[height=3.2 in,width=2.7in,angle=-90]{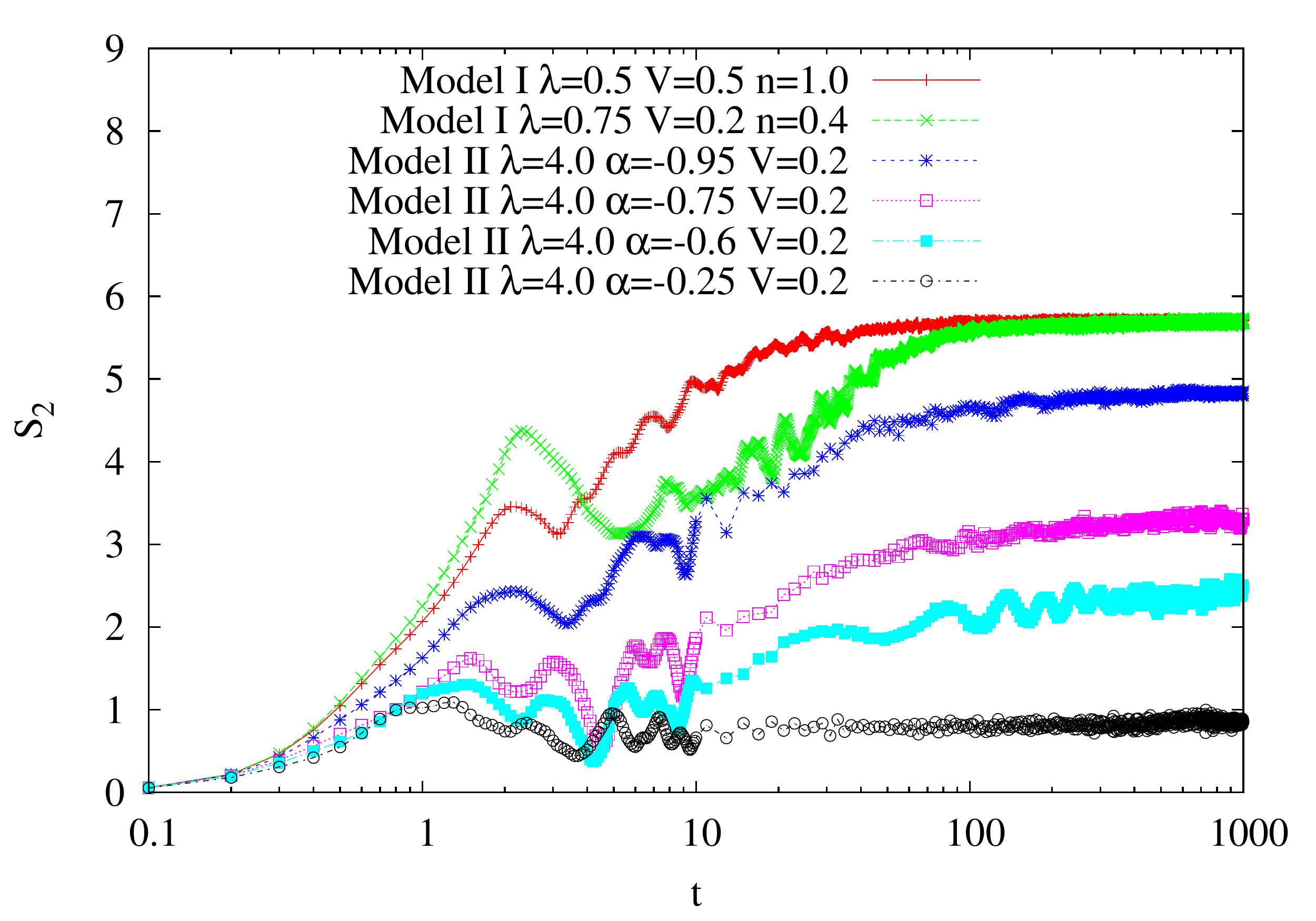}
\caption{Variation of the Renyi entropy for $L=14$ at half filling for the two models 
with different parameters. Figure is taken from Ref.~\onlinecite{modak2015many}.}
\label{Fig:entropy II}
\end{figure}

\begin{figure}
\begin{center}
\setlength{\unitlength}{8.5cm}
\begin{picture}(1.05, 0.818)(0,0)
   \put(0,0.8){\resizebox{1\unitlength}{!}{\includegraphics[angle=-90]{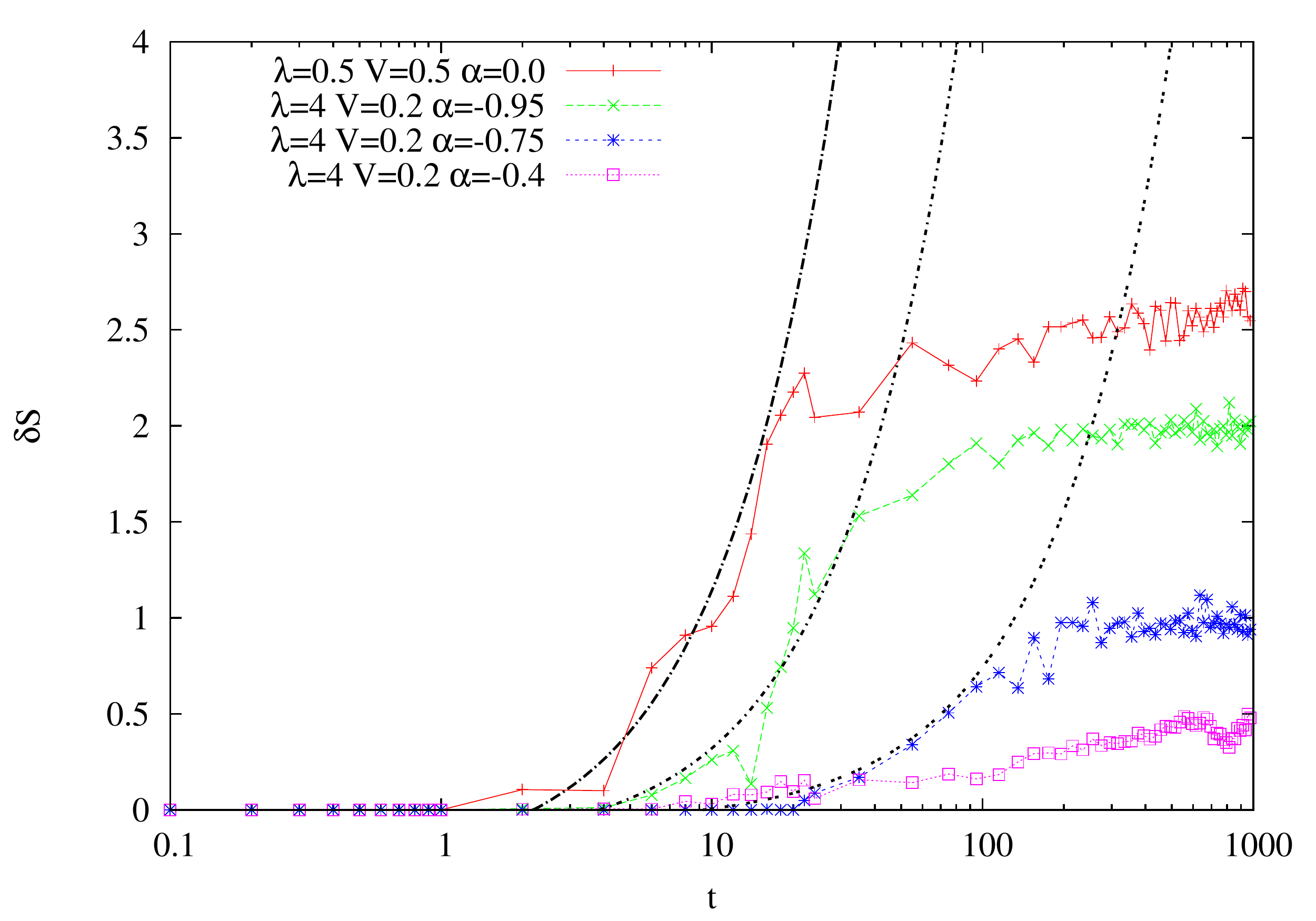}}}
\put(0.15,0.63){\resizebox{0.41\unitlength}{!}{\includegraphics[angle=-90]{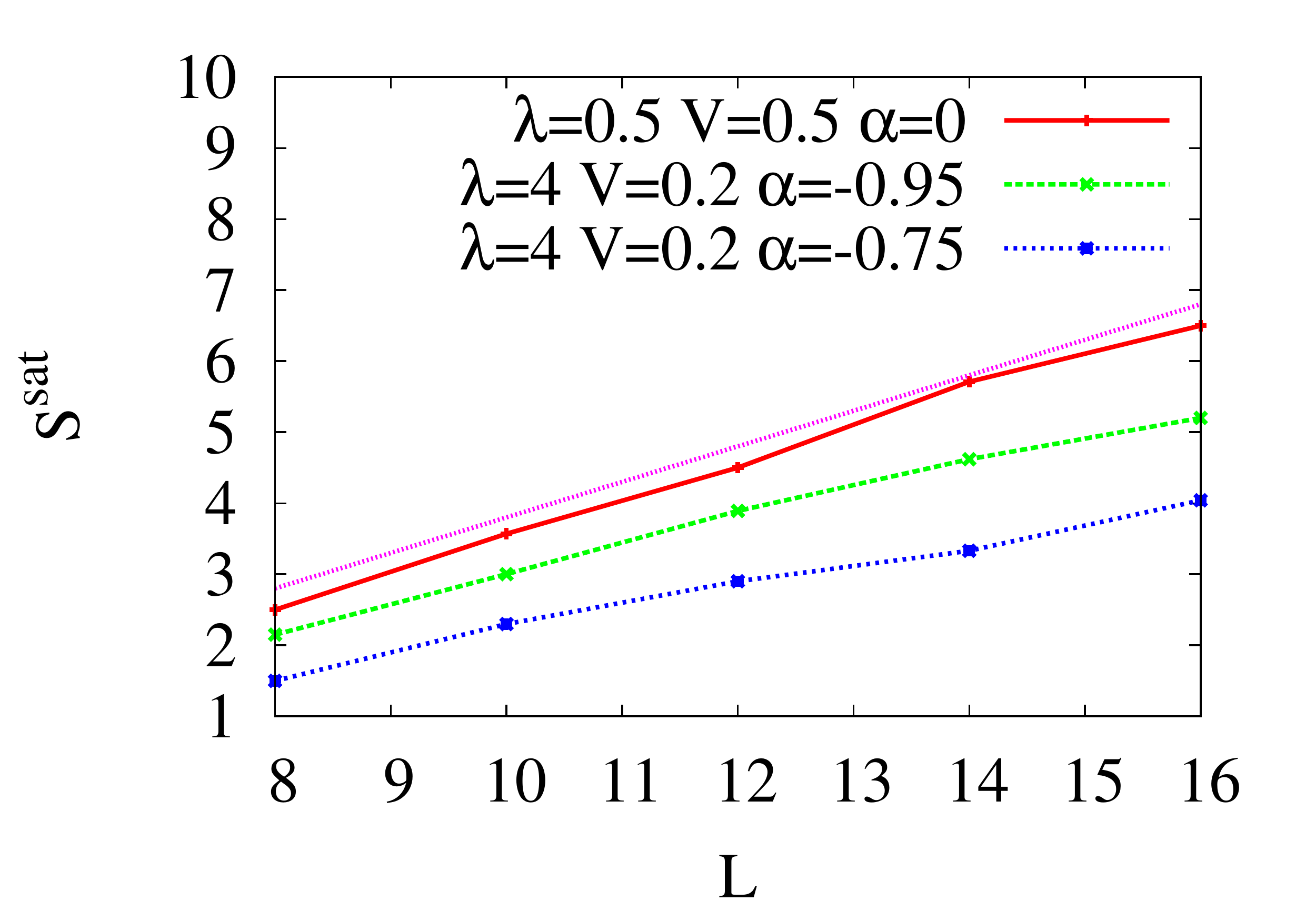}}}
\end{picture}
\end{center}
\vspace{-0.6in}
\caption{The variation of $\delta S=S_2(t,V)-S_2(t,V=0)$. $S_2$ is the Renyi entropy for $L=14$ at half filling for 
the two models with different parameters. The dotted lines are linear fits in $t$. 
(Inset) The variation of the saturation value of $S_2$ with $L$. The blue dotted line corresponds to 
thermal value of $S_2=\frac{L}{2}-1.2$ for system size $L$. Figure is taken from Ref.~\onlinecite{modak2015many}}
\label{Fig:delta S}
\end{figure}

As shown in Fig.~\ref{Fig:entropy II}, $S_2(t)$ for model I increases linearly with time and  then  saturates to $L/2$. In contrast, for model II, $S_2$  saturates to a  
much smaller value than $L/2$, which is a signature of the MBL phase.
However, the growth of entanglement in model II appears to be much faster than logarithimic as shown in Fig.~\ref{Fig:delta S}.
Thus, from the growth of entanglement, it appears that model I is ergodic when interactions are switched on while model II is localized consistent 
with the conclusions arrived at from the level statistics. However, it is curious that the initial growth of entanglement in model II is linear and not logarithmic as one would have expected for an MBL phase. Since we are considering the \emph{energy averaged} entanglement entropy, it is natural to associate the initial linear growth with the existence of extended states and the presence of a many-body mobility edge. We explore this further in the following subsection.


\subsubsection{Energy resolved entanglement scaling and the violation of ETH}
\begin{figure*}
\centering
\includegraphics[scale=0.65]{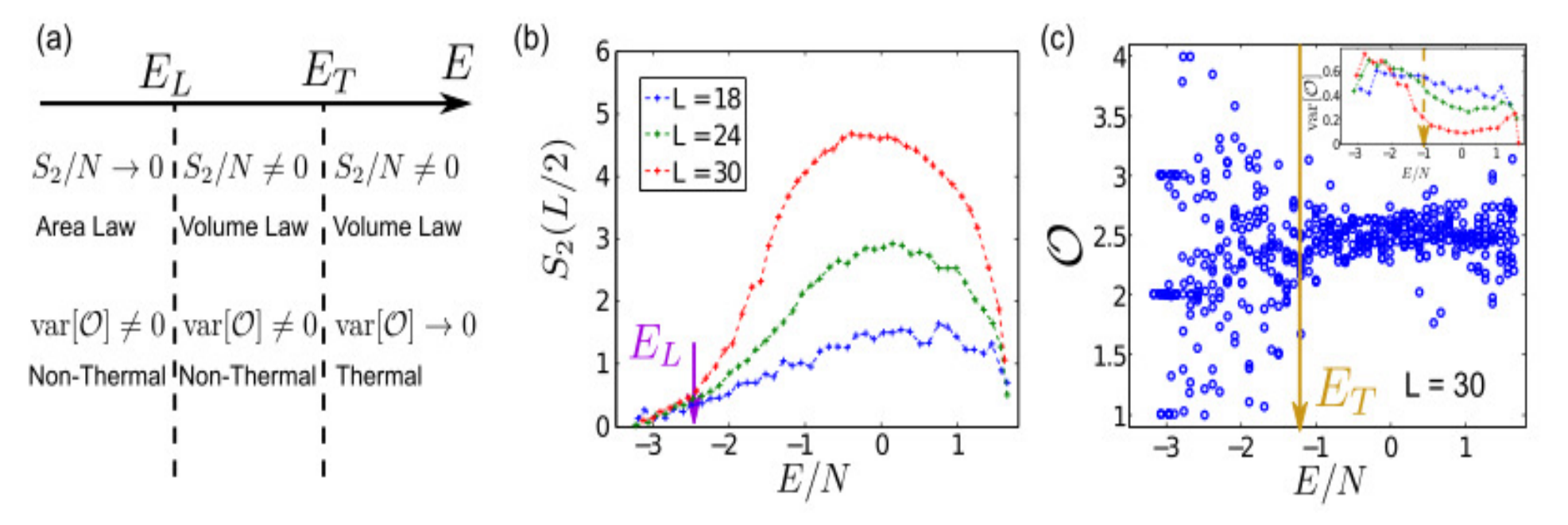}
\caption{Localization and ergodicity in model II as determined from entanglement scaling and verification of ETH. (a) Schematic showing the three different phases as a function of energy. The eigenstates with $E < E_L$ are localized, the ones with $E_L < E_T$ are extended but non-ergodic and the ones with $E > E_T$ are ergodic. (b) The entanglement scaling showing the location of $E_L$  . (c) The expectation value of the operator $O$, which is the density of particles in one half of the system. $E_T$ is obtained from the size of the fluctuations of this quantity, which can be quantified by the variance (inset). The numerical data for the middle and right panels were obtained from systems of size $L=30$ and $5$ particles. The figure is taken from Ref.~\onlinecite{li2015many}.}
\label{Fig:li_et_al}
\end{figure*}

MBL is believed to be concurrent with non-ergodicity. The signature of localization is an area-law scaling of the entanglement entropy in an energy eigenstate whereas that of non-ergodicity is a violation of ETH. $S_2$ has been calculated for model II as a function of energy density and it has been found that there is a many-body mobility edge of energy density $E_L$ as shown in Fig.~\ref{Fig:li_et_al} (Ref.~\onlinecite{li2015many}). Eigenstates with energy density $E < E_L$  are many-body localized whereas those with $E >E_L$ are extended. The validity of ETH for this model has also been examined and it has been found that there is a threshold of energy density $E_T$ such that for states with $E < E_T$, ETH is violated but for those with $E > E_T$, it is not. Interestingly $E_L \neq E_T$ and in fact $E_L < E_T$ so that states with $E_L < E < E_T$ are extended but non-ergodic. Thus, there is an energy window in which the system behaves as a non-ergodic metal. 

A picture of the three different kinds of many-body states, (i) localized and non-ergodic; (ii) extended and non-ergodic; (iii) extended and ergodic can be obtained by thinking in terms of the many-body states of the non-interacting system. The single-particle states of the non-interacting system are either localized or delocalized. Many-body states can be constructed by populating only the localized states which are many-body states of type (i), both localized and delocalized states, which are of type (ii) and finally only delocalized states, which are of type (iii). Note, that in the absence of interactions, all three types of states are non-ergodic and type (iii) becomes ergodic only upon turning on interactions.

The existence of a many-body mobility edge and in particular, extended non-ergodic states is consistent with the observation from entanglement growth that the $S_2(t)$ increases roughly linearly in time (as is expected for an extended system) but saturates to a sub-thermal extensive value (as is expected for a non-ergodic system).

\subsubsection{Criterion for non-ergodicity}
The studies of Refs.~\onlinecite{li2015many,modak2015many} suggested that 
a non-ergodic phase can exist in a system with  mobility edges (where 
 localized and  delocalized states are not protected)
upon the introduction of weak interactions. Model II (as described in the 
previous section) is an example of such system. However, the model studied (model I)
appears to thermalize which raises the question of how to distinguish between models with single-particle mobility edges 
that will thermalize or remain non-ergodic upon the introduction of interactions~\cite{modak2015many}. More precisely, 
can a prediction 
be made about the behavior of the model upon the introduction of weak interactions based on the spectrum without interactions?
Very recently, there has been an attempt to answer this question by proposing a criterion~\cite{modak2016criterion} based 
on the value of a single parameter ($\epsilon$) obtained from the single-particle spectrum. This criterion is empirical and has been verified by 
extensive numerical  studies on several different models with single-particle mobility edges. The quantity $\epsilon$ is defined as   
\begin{equation}
\epsilon= \frac{\eta(1-\frac{MPR_D}{L})}{(MPR_L -1)},
\label{Eq:defratio}
\end{equation}
and is a measure of the the relative strengths
of the localized and delocalized states in the non-interacting model, i.e. how strongly localized the localized states
are compared to how strongly delocalized the delocalized ones are. 
$\eta$ is the ratio of the number of localized states to delocalized single-particle orbitals,
$MPR_D$ ($MPR_L$) is the mean participation ratio of the delocalized (localized) states.
The proposed criterion is that the system remains localized 
(thermalizes) upon the introduction of weak interactions when $\epsilon > (<1)$. 
 It is important to emphasize that the criterion is `only heuristic' and a better theoretical 
understanding is required for its justification.

\subsection{The MBL proximity effect}
 The numerical studies described in the previous section providing evidence for the lack of ergodicity in systems with single-particle mobility edges have motivated 
 further investigations of the coupling of localized and delocalized bands of states by interactions.  
 Nandkishore~\cite{nandkishore2015many} has considered a system where the back-action of a localized interacting system on a bath of delocalized states 
 cannot be neglected thereby offering no protection against localization. It emerges that the MBL of the system not only survives but it also causes the bath to get localized due to the coupling.
    
 Specifically, the system and the bath are thought to consist of two species of spinless fermions of the $c$ and $d$ type respectively on 
 a $D$-dimensional lattice. The system is described by the following Hamiltonian.
 \begin{eqnarray}
  H_c=\sum_i\mu_i c_i^{\dag}c_i+\sum_{\langle i, j \rangle}t_c (c_i^{\dag}c_j +\mathrm{h.c.}) +Uc_i^{\dag}c_ic_j^{\dag}c_j \,\,\,\,
 \end{eqnarray}
  where, $\langle i, j \rangle$ denotes a sum over nearest neighbors (NNs), with a hopping $t_c$, and interaction $U$. $\mu_i$ is a
  random on-site potential, drawn independently from a even distribution of width $W$. 
   The width is chosen to be sufficiently large such that 
  the $c$ fermions  are in MBL phase with a localization length $\xi_c$ and it is assumed that $W$ is the largest energy scale 
  in the problem.
  The ergodic bath is described by the following Hamiltonian for d type fermions. 
  
\begin{eqnarray}   
  H_d=\sum_{\langle i,j \rangle}t_d (d_i^{\dag}d_j +\mathrm{h.c.}) +\lambda d_i^{\dag}d_id_j^{\dag}d_j 
 \end{eqnarray}
  where, $t_d$ is the NN hopping and $\lambda$ is the NN interaction between $d$ type fermions.
 The coupling between the system and the bath ($c$ and $d$ type fermions) was taken to have the form
 \begin{eqnarray}
  H_{int}=\sum_{i}g_i c_i^{\dag}c_id_i^{\dag}d_i 
   \end{eqnarray}
and the $g_i$'s are taken independent from a even distribution of width $G$. 

\subsubsection{The limit $t_d<G<W$} 
At $t_d \to 0$  the effective random potential seen by the $c$ fermions is 
be increased from $W$ to $\sqrt{W^2+G^2}$ due to the distribution of the $d$ fermions. Hence, if the $c$ fermions are localized at $G=0$, they remain localized for non-zero
$G$.  For $t_d \neq 0$, the hopping of a $d$ fermion will take the system off shell in energy by   $\Delta E\sim W\exp(-s(T)\xi_c^D)$ 
($s(T)$ is the entropy density). By comparing the change of the energy with hopping matrix elements between the two states, it can be argued that locator expansion~\cite{anderson1958absence} converges if 
the typical hopping matrix element is smaller than the amount of change in energy by which the $d$ fermion hopping takes 
the system off shell. This provides an upper bound on $t_d$. A calculation shows that the locator expansion converges if  $t_d<\min \left\lbrace G,W \exp(-\frac{1}{2}s(T)\xi_c^D)\right\rbrace$ 
and hence, in that limit the combined $c$ and $d$ systems are localized. 

\subsubsection{The limit $\lambda<<G<<t_d<<W$}
 In the limit $\lambda \to 0$, the $d$ fermions are described by the following Hamiltonian.
 \begin{eqnarray}
  H_d=\sum_{<ij>}t_d d^{\dag}_id_j +h.c. +\sum_i V_i d^{\dag}_i d_i
 \end{eqnarray}
where, $V_i=g_ic_i^{\dag}c_i$ is the effective disorder potential seen by a $d$ fermion and the precise disorder realization
depends on the state in which the $c$ fermions are prepared. In $D=1$ and 2 an arbitrarily small amount of disorder is sufficient to localize all the states
with a localization length $\xi_d$ which is a power law large in $t_d/G$ for $D=2$ and exponentially large in $t_d/G$ for $D=1$ (Refs.~\onlinecite{anderson1958absence,abrahams1979scaling}). In the limit $\xi_c\to 0$ 
the $c-d$ coupling is diagonal in the $c$ eigenbasis and the back action on the $c$ system is not appreciable. Thus, the whole system localizes.
For nonzero $\xi_c$, there is an effective four $d$ fermion
 interaction mediated by the $c$ fermions and the locator expansion converges if the matrix elements of the interaction are less than the level spacing . 
It can be shown that a leading order perturbation theory calculation in weak $G$  gives rise to a criterion for the convergence of the locator expansion, i.e.
\begin{eqnarray}
 \max\left\lbrace\frac{G^2}{Wt_d\xi_d^{D/2}}\exp(-2/\xi_c),\lambda/t_d\right\rbrace\xi_d^{3D} <1;  \xi_c<1 \nonumber \\
  \max\left\lbrace\frac{G^2 \xi_c^{2D}}{Wt_d\xi_d^{D/2}},\lambda/t_d\right\rbrace\xi_d^{3D} <1;  \xi_c>1
\end{eqnarray}

\subsubsection{The limit $G\to \infty$}
For any arbitrary $\lambda$, $t$ and in the limit $G\to \infty$, 
there are three
types of states of the fermions, (1) unbound $c$ fermions, (2) bound states where the $c$ and $d$ fermions
sit on the same site and (3) unbound $d$ fermions and a bound state cannot be broken apart or form because costs energy of the order of $G$.
Since, the unbound $c$ fermions are governed by a Hamiltonian such that they are already localized,  some lattice sites are forbidden due to the presence of bound $c-d$ pairs. 
This gives rise to enhanced obstruction in transport and hence, unbound $c$ fermions get localized even more strongly. The bound $c-d$ pairs are very heavy and the effective hopping matrix elements are of the the order of $t_ct_d/G<<t_c$. Thus, the effective random potential seen by the $c-d$ pairs is the same as the $c$ fermions.  Since, the $c$ fermions are localized, these pairs must also be localized. The effective lattice for the unbound $d$ fermions is obtained from the original lattice by removing all the sites on 
which unpaired $c$ fermions and $c-d$ pairs are present. In $D=1$ even one deleted site creates an obstruction of transport. Hence, the $d$ fermions are also localized 
and the many-body proximity effect occurs. 
\begin{figure}
\includegraphics[width=3in]{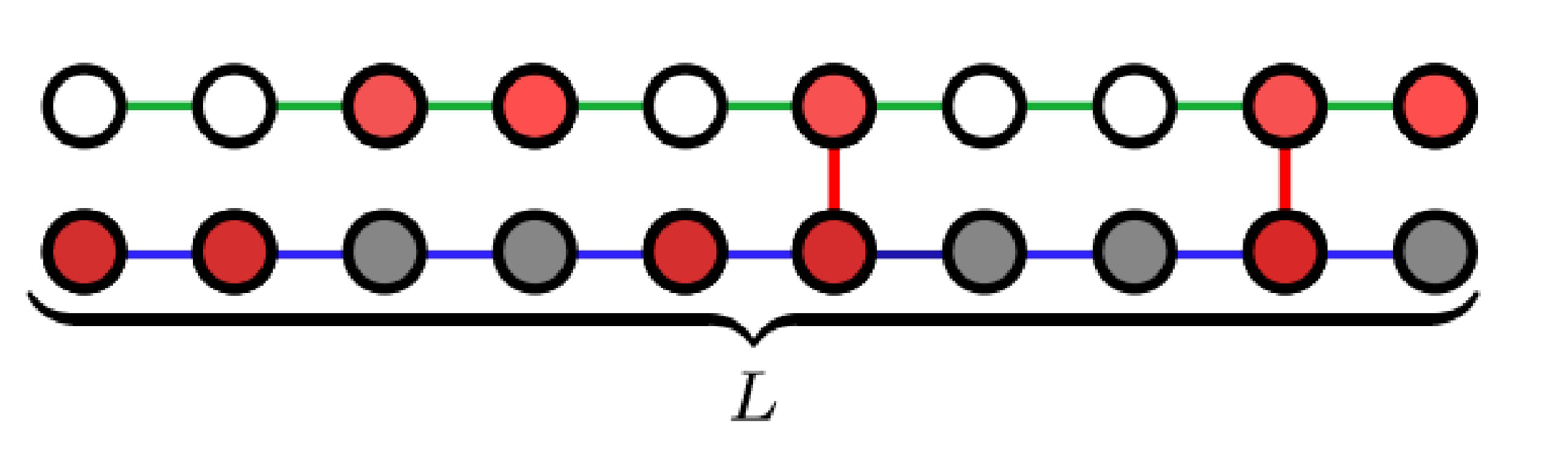} 
\caption{ Schematic of the model on a ladder, where  the disorder
potential only acts on the lower chain of the ladder  and  the fermions in the upper chain are affected by
the disorder only through the interactions. The fermions,
indicated as red dots, hop along each chain, and interact only
through a  density-density interaction (Figure taken from ~\cite{hyatt2016many})}
\label{Fig: lattice}
\end{figure}

\subsection{Numerical evidence for the proximity effect}
A recent numerical study by Hyatt et al. (in Ref.~\onlinecite{hyatt2016many}) has found evidence for the many-body proximity effect. 
They have studied a system of spinless fermions on a ladder, where one chain of the ladder is translationally invariant and the 
other experiences a disorder potential with a density-density interaction between the two chains. 
The  system is described by the following Hamiltonian.
\begin{eqnarray}
 H=-\sum_{\kappa}t_{\kappa}\sum_{i=1}^{L}(c_{\kappa,i}^{\dag}c_{\kappa ,i+1} +h.c.) \nonumber \\
 +\sum_{i=1}^{L}w_in_{d,i}+V\sum_{i=1}^{L}n_{d,i}n_{c,i}
\end{eqnarray}
where, $c_{\kappa,i}^{\dag},c_{\kappa ,i}$ are creation annihilation operator on the clean chain (upper chain according to  Fig.~\ref{Fig: lattice}) when $\kappa=c$ and 
on the disordered  chain (lower chain according to Fig.~\ref{Fig: lattice}) when $\kappa=d$. $w_i$, the on-site random potential on the lower chain, is drawn 
from a uniform random distribution in the interval between $[-W,W]$. $V$ is the density-density interaction between two chains and $L$ is the length
of each chain. 

In this model, in the limit $V\to 0$  localized and delocalized degrees of freedom coexist.  Hence, this model allows one to study the effect of coupling
between a small bath and a many-body localized system. Here, the upper clean chain can be considered as a  bath with delocalized degrees of freedom and the lower chain
can be considered as a localized system. It is found that in the presence of strong disorder $W=8$, a weakly delocalized bath ($t_c<0.2$)(keeping $t_d=1)$ and moderate interactions, an MBL phase can exist as suggested by different diagnostics such as the time evolution of the entanglement entropy and scaling of the eigenstate entanglement entropy. However, since, the parameter space is mainly dominated by the delocalized phase, the authors are not able to rule out the possibility of delocalization on very long length scales.

\subsection{MBL due to random interactions}
Very recently,  the existence of MBL has been studied with in models that have a complete absence of localized orbitals. In particular, both the SU(2) symmetric Hubbard model~\cite{2016_Lev_arXiv} and spinless fermion models~\cite{2016_Li_MBL_arXiv} have been considered.  Using exact diagonalization and density-matrix-renormalization-group methods, it has been established that the mobile fermions get localized in the presence of random interactions. While MBL essentially arises from the ``continuous deformation'' of  Anderson localization in the presence of interactions in the conventional paradigm,  it has been argued that the random-interaction induced localization, ``dubbed" statistical bubble localization, goes beyond. For example, the LIOM description may not apply to the bubble localization, due to the existence of rare long-range entangled many-body states~\cite{2016_Li_MBL_arXiv}.  

\section{Experiments}
In this section, we discuss the recent experimental progress of single-particle localization and MBL in incommensurate systems. Without interactions, experimental observations of Anderson localization has been reported in various setups, for example scattering light in GaAs (Ref.~\onlinecite{Wiersma-1997-Nature}) or $\text{TiQ}_2$ (Ref.~\onlinecite{Sperling-2013-NaturePhotonics}) powders and photonic lattices \cite{Segev-2013-NaturePhotonics,Schwartz-2007-Nature,Yoav-2008-PRL,Lahini-2009-Observation}, cold atoms with random \cite{Kondov-2011-Science,Billy-2008-Nature,Jendrzejewski-2012-NaturePhysics} or quasirandom \cite{Roati-2008-Nature} disorder
potentials, 
 ultrasound in an elastic network \cite{Hu-2008-NaturePhysics}, entangled photons in integrated quantum walk \cite{Crespi-2013-NaturePhotonics}, and electrons in single crystals  $\text{Li}_x\text{Fe}_7\text{Se}_8$ (Ref.~\onlinecite{Ying-2016-ScienceAdvances}). In particular, in Ref.~\onlinecite{Lahini-2009-Observation}  the signature of localization of light was observed in 1D quasiperiodic photonic lattices that realized the AA model. In this setup, the localization transition was obtained by  directly measuring the spread of the initial narrow wave packets. 
Below the predicted transition point, the initial narrow wave packet spreads out as it propagates in the lattice, whereas
above the transition the expansion was drastically suppressed, a clear indication of localization.  For ultracold atoms in incommensurate optical lattices, Anderson localization has been reported in Ref.~\onlinecite{Roati-2008-Nature}, where the localization transition was observed by measuring transport properties and spatial and momentum distributions of a 
Bose-Einstein condensate.
The scaling behaviour of the critical disorder strength was also studied experimentally.

The experimental observation of MBL is expected to be more elusive and has only been reported very recently in cold atoms \cite{Schreiber-2015-Science, Choi-2016-Science,Bordia-2016-PRL,Luschen-2016-arXiv,Bordia-2016-arXiv,Marcuzzi-2016-arXiv} and trapped ions \cite{Smith-2016-NaturePhysics,Zhang-2016-arXiv}, which are both very close to a perfectly isolated quantum system. Particularly, in Ref.~\onlinecite{Schreiber-2015-Science} the Bloch group has reported the observation of MBL of ultracold fermions in a 1D quasi-random optical lattice. By superimposing an incommensurate lattice on the primary 1D lattice and tuning a magnetic Feshbach resonance \cite{Chin-2010-RMP}, they 
have realized
the interacting AA model. 
They prepared  an initial high energy charge density wave (CDW) state, and the atoms only occupied the even lattice sites only.
They then monitored the time evolution of the initial CDW under different parameter regimes and measured the time-dependent imbalance by the band-mapping technique. They observed that, while for zero or weak disorder the stationary value of the imbalance approaches zero, for stronger disorder it remains finite for all observation times, indicating a breaking of ergodicity  and a signature of MBL.  In Ref.~\onlinecite{Bordia-2016-PRL}, they moved one step further to couple an array of identical 1D incommensurate lattices. In this case, they found that, in the absence of interactions the coupled system  remained localized, but even very weak couplings would delocalize the MBL phase with interaction.  This manifests an intriguing difference between MBL and Anderson localization. In Ref.~\onlinecite{Bordia-2016-arXiv}, the Bloch group has experimentally studied the localization features of the same 1D incommensurate lattices with a periodic driving. They found two distinct phases, one localized and the other ergodic, separated by a dynamical phase transition that depends on both the drive frequency and drive strength. It is also worthwhile to mention that signatures of MBL has also been observed in an 1D open fermionic system with controlled dissipation \cite{Luschen-2016-arXiv} and in a 2D randomly disordered optical lattice with strongly interacting bosons \cite{Choi-2016-Science}.

\subsubsection{Optical lattice realization of the GAA model}
One of the advantages of working with the GAA model is that it can be realized with ultracold atoms in optical lattices. We  showed that a limit of the GAA model can be accessible in the optical lattices within existing experimental technology. In the limit of $\alpha\ll 1$, the on-site potential can be approximated by,
\begin{equation}
	\frac{\cos(2\pi n b+\phi)}{1-\alpha \cos(2\pi n b+\phi)}\sim\cos(2\pi n b+\phi)+\alpha \cos^2(2\pi n b+\phi).
	\label{eq:gaaexp}
\end{equation}
\begin{figure}
\centering
\includegraphics[scale=0.52]{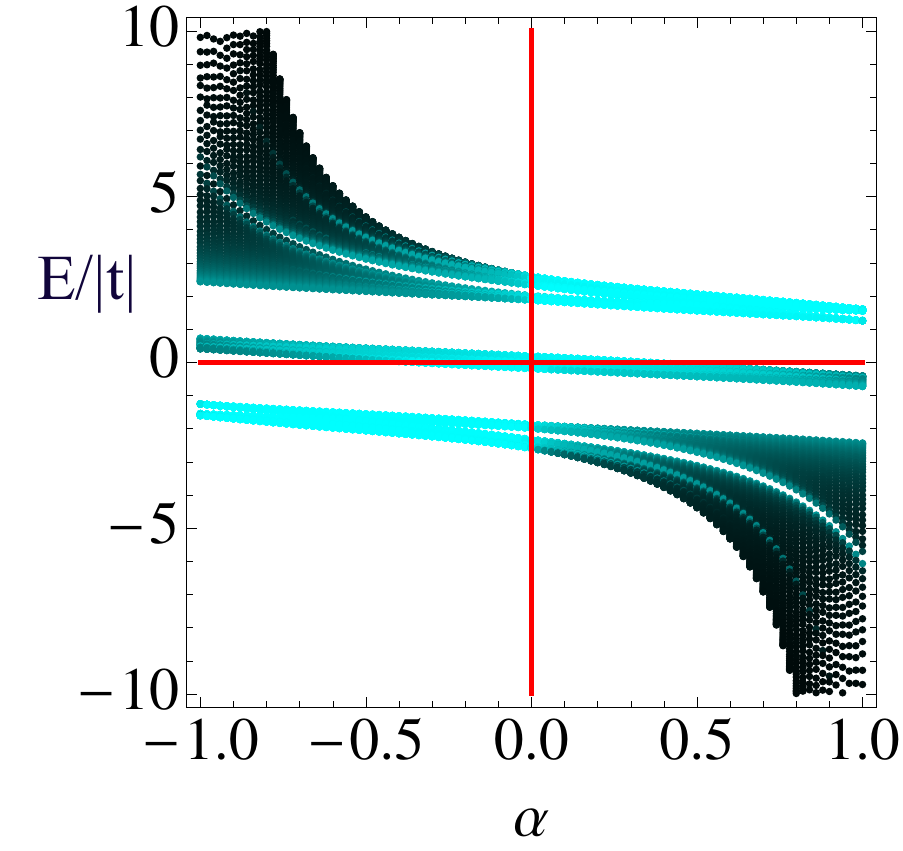}\\
\includegraphics[scale=0.5]{figlegend.pdf}
\caption{\label{gaaexp} Localization transition in the GAA model at $|\lambda|=|t|$ point. Spectrum of GAA model as a function of $\alpha$ with the red line corresponding to $\alpha E=0$ critical point. Black corresponds to the IPR~1 corresponding to the fully localized state. Cyan denotes the value $1/L$ (where $L=200$ is the number of sites) corresponding to the fully extended state.}
\end{figure}
Notice that the first term is the same as the AA model that has been well studied experimentally. The second term can be realized by adding another laser that is half the wavelength of the second incommensurate laser. Note that for the choice of $|\lambda|\sim |t|$ (the AA critical point), the critical line reduces to $\alpha E\sim 0$. 
For any small non-zero $\alpha$, the critical energy line is $E=0$. All the states corresponding to $E>0$ are localized and all the states for $E<0$ are extended as shown in Fig.~(\ref{gaaexp}). The thermalization or lack thereof in the interacting GAA model can be probed following Ref.~\onlinecite{Schreiber-2015-Science}. The AA model in the above MBL experiment can be deformed into the GAA model defined in Eq.~(\ref{eq:gaaexp}) by using the laser that was previously used for creating the inhomogeneous CDW profile. 

\section{Summary and conclusions}

To summarize, in this article we have reviewed the current state of research on MBL in models with incommensurate potentials in 1D. In the absence of interactions in 1D, such models can possess single-particle delocalized states in contrast to those with random potentials which have only localized states. While the most commonly studied incommensurate model, the AA model has two phases, one with all single-particle states localized and another with  all delocalized, generalized incommensurate models have both types of single-particle states 
which are 
separated by a mobility edge. LIOM can be constructed for these non-interacting models as power series in the hopping, 
whose validity relies on the convergence.
 The number of such LIOM is 
 found to be 
equal to the number of localized states. 

Three different types of many-body states of fermions can be constructed for these non-interacting models by selectively populating (i) only delocalized states, (ii) only localized states and (iii) partial delocalized and partial localized states. The states in category (i) obey ETH and show a volume-law scaling of the entanglement entropy, which is also equal to the thermodynamic entropy. The category (ii) states violate ETH and show an area law scaling of the entanglement entropy. The states in category (iii) violate ETH but show a volume-law scaling of the entanglement entropy. However, this entropy is less than the thermodynamic entropy that the subsystem would possess if it were to thermalize. 

In the absence of interactions, an incommensurate model of the sort described above does not thermalize in the sense that 
an arbitrary initial state of the system does not necessarily yield a thermal density matrix for subsystems at long (or infinite) time limit.
Interactions give rise to non-zero matrix elements among all of the different types of many-body states described in the previous paragraph which determine whether the system thermalizes or not. It has been argued that systems with single-particle delocalized states and ``protected'' delocalized states generically thermalize upon the introduction of arbitrarily weak interactions when the localization length of the localized states diverges with exponent $\nu > 1$ upon approaching the mobility edge (in 1D). For the marginal case $\nu=1$, the system does not thermalize if the matrix elements for ``flip-flop'' assisted hopping fall off sufficiently slowly. For models with incommensurate potentials, there is {\em no} protection for the delocalized states. Nevertheless, it emerges that the matrix elements mentioned above fall off rapidly for those models which do not thermalize upon the introduction of interactions and not for those that do. 

Incommensurate models display non-ergodicity upon the introduction of interactions despite the presence of delocalized states. This can be seen by numerically obtaining the many-body energy eigenstates and eigenvalues for specific microscopic models. The level spacing statistics, entanglement growth and saturation, eigenstate entanglement scaling and tests of ETH are consistent with lack of ergodicity. As a function of energy, the many-body eigenstates of the interacting system are also of the types (i), (ii) and (iii) described above for the non-interacting system and are separated by appropriate``many-body mobility edges''. In particular, the states in category (iii) describe a non-ergodic delocalized (metallic) phase, where the system does not thermalize but the entanglement entropy of the energy eigenstate 
obeys a volume law scaling, yet with a sub-thermal value.
This is also consistent with the observed behavior of the entanglement growth starting from an ensemble of equally weighted unentangled states at all allowed energies.  

Not all incommensurate models are non-ergodic upon adding interaction. Whether a given model thermalizes or not appears to depend on the nature of the localized and delocalized single-particle states, specifically how strongly localized or delocalized they are as characterized by their IPR. A heuristic criterion in terms of a ratio of the weighted IPR values of the two kinds of state can be formulated in order to guide 
whether a given model will thermalize or not upon the introduction of interactions. The lack of ergodicity in an incommensurate model is a consequence of the fact that when interactions are introduced, the localized states are localized strongly enough to introduce non-ergodicity in the whole system, i.e. the localized states localize the delocalized states.  This goes by the name of the many-body proximity effect and exists even in systems with random potentials (and hence localized single-particle states) that are coupled to other systems with delocalized states as revealed by analytical arguments and numerical studies. It has also been shown recently that localized single-particle states are not required to produce MBL 
one example being delocalized fermions subjected to random interactions.

Finally, incommensurate models have been realized in experiments. In fact, the first experimental observation of MBL was in a cold-atomic system 
which is described by
the AA model with interactions. This system does not possess a single-particle mobility edge but a suitable modification of the experimental apparatus with an additional laser to generate a second harmonic can produce a system with a 
controllable
single-particle mobility edge. Thus, the existence of non-ergodicity in the presence of a single-particle-mobility edge and the non-ergodic metallic phase can be tested experimentally.

\begin{acknowledgements}
We thank Sankar Das Sarma for useful discussions and collaborations on related work. This work has been partially supported by JQI-NSF-PFC, ARO-Atomtronics-MURI, LPS-MPO-CMTC, and Microsoft Q (DLD, XL, and JHP). DLD and JHP also acknowledge additional support from the PFC seed grant ``Thermalization and its breakdown in isolated quantum systems''. SM thanks the ISF-UGC program for support.
\end{acknowledgements}

\end{document}